\documentclass[letterpaper,12pt]{article}
\usepackage{latexsym,amssymb,amsmath, bm}
\usepackage{amsthm}
\usepackage{relsize}
\usepackage{lipsum,graphicx,multicol}
\usepackage[colorlinks=true,allcolors=blue, linkcolor=red]{hyperref}
\usepackage{setspace,color}
\usepackage[sort&compress]{natbib}
\usepackage{tcolorbox}
\usepackage{float}
\usepackage[normalem]{ulem}
\usepackage{authblk}
\usepackage{leftidx}
\usepackage{dsfont}
\usepackage{geometry}
\usepackage[title,titletoc]{appendix}
\usepackage{actuarialangle}
\usepackage{pdfpages}
\usepackage{caption}
\usepackage{subcaption}
\usepackage{mathtools}
\usepackage{makecell}
\allowdisplaybreaks

\usepackage{lineno}
\usepackage{bbm}

\usepackage{accents}

  \renewenvironment{abstract}
  {\par\begin{center}\bfseries Abstract\end{center}\ignorespaces}
  {\par}

\def \P {\mathbb{P}}

\def \uu {\bm{u}}

\def \XX {\bm{X}}
\def \xx {\bm{x}}

\def \one {\bm{1}}

\def \d {\textrm{d}}

\def \cc {\boldsymbol{c}}

\def \d {\mathrm{d}}
\def \ee {\boldsymbol{e}}

\def \gg {\boldsymbol{g}}

\def \uu {\boldsymbol{u}}

\def \xx {\boldsymbol{x}}

\def \A {\mathcal{A}}
\def \B {\mathcal{B}}
\def \P {\mathcal{P}}
\def \AA {\mathfrak{A}}
\def \BB {\mathfrak{B}}

\def \C {\mathrm{C}}

\def \GG {\boldsymbol{G}}

\def \PP {\mathbb{P}}

\def \RR {\boldsymbol{R}}
\def \R {\mathbb{R}}

\def \SS {\mathfrak{S}}

\def \UU {\boldsymbol{U}}
\def \VV {\boldsymbol{V}}
\def \WW {\boldsymbol{W}}
\def \XX {\boldsymbol{X}}
\def \YY {\boldsymbol{Y}}
\def \ZZ {\boldsymbol{Z}}

\def\e{{\mathbb E}}

\def\Cov{{\mathrm{Cov}}}
\def\Var{{\mathrm{Var}}}

\def\ac{{\mathrm{ac}}}
\def\co{{\mathrm{co}}}

\newtheorem{theorem}{\bf Theorem}

\newtheorem{example}{\bf Example}
\newtheorem{remark}{\bf Remark}
\newtheorem{proposition}{\bf Proposition}
\newtheorem{lemma}{\bf Lemma}
\newtheorem{corollary}{\bf Corollary}


\definecolor{darkread}{rgb}{0.7, 0, 0}

\newcommand{\SF}{\,\overline{\kern-0.175em F\kern-0em}}

\newcommand{\sol}[1]{%
  \mkern 1mu\overline{\mkern-1mu#1\mkern-0.25mu}\mkern 1mu
}

\begin{document}

\begingroup
\onehalfspacing
\renewcommand{\thefootnote}{\fnsymbol{footnote}}

\begin{center}
  {\LARGE\bfseries
  Dirichlet sub-mixtures: A new reading of\\
  multiplicative background risk models with Pareto margins\par}

  \vspace{0.6cm}
  {\large Nawaf Mohammed\footnotemark{} and Edward Furman\par}

  \vspace{0.4cm}
  Department of Mathematics and Statistics, York University,
  Toronto, ON M3J 1P3, Canada \\
  RISC Foundation, 5000 Yonge Street, Suite 1901, Toronto, ON M2N 7E9, Canada
\end{center}

\footnotetext{%
  Corresponding author. E-mail address: \href{mailto:nawaf@yorku.ca}{%
    \textcolor{blue}{\texttt{nawaf@yorku.ca}}}}

\vspace{0.2cm}

\begin{abstract}
Multiplicative background risk models (MBRMs) provide a tractable framework for describing dependent risks. We construct a class of MBRMs with Pareto margins by combining Marshall--Olkin exponential and additive gamma risk factor structures, each allowed to incorporate idiosyncratic and systemic background risk factors. Our main result establishes, under an admissibility condition on the risk factor systems, a finite sub-mixture representation of the joint survival function in terms of Pareto survival functions, whose weights are Dirichlet probabilities of orderings induced by the background risk factors and sum to at most one. This yields a new interpretation of such MBRMs, in which dependence is expressed through Pareto survival functions and their Dirichlet weights. The resulting family encompasses many known multivariate Pareto constructions, is closed under marginalization, and admits characterizations of independence and, under a non-degeneracy condition, of its singular and absolutely continuous parts. It is associated, its covariance decomposition separates the contributions of the two risk factor structures and their interaction, and its joint tails are regularly varying along rays. The presence of asymptotic tail dependence and the decay rates of joint exceedance probabilities are governed by the gamma risk factors alone; the exponential risk factors affect only the tail constants.


\end{abstract}

\par
\endgroup

\clearpage
\setcounter{footnote}{0}

\newpage
\section{Introduction}
\label{sec:introd}

Heavy-tailed phenomena arise naturally in insurance and finance, where a handful of extreme risks can dominate the aggregate outcome of a collection of risks \citep{Embrechts1997}. Among the distributions used to model such phenomena, the Pareto distribution \citep{Pareto1964} occupies a particularly prominent position owing to both its analytical tractability and its fundamental role in extreme-value theory. Indeed, the celebrated theorem of \cite{Balkema1974} and \cite{PickandsIII1975} establishes that exceedances over sufficiently high thresholds are approximately (generalized) Pareto distributed. More precisely, if the risk random variable $Y$ has a survival function that is regularly varying at infinity with index $-\alpha$, $\alpha>0$, then the excess random variable $X=(Y-u)\mid Y>u$ at the threshold $u>0$ and with survival function $\SF_X$ satisfies
\begin{equation*}
\lim_{u\to\infty}\,\sup_{x\ge 0}
\left|\SF_X(x)-\left(1+\beta(u)x\right)^{-\alpha}\right|=0
\end{equation*}
for a suitable rate function $\beta(u)>0$, e.g., $\beta(u)=1/u$; see \cite{Embrechts1997} for further details. The Pareto distribution of the second kind, also known as the Lomax distribution, with survival function $(1+\beta x)^{-\alpha}$, $x\ge0$, is therefore a canonical model for heavy-tailed insurance risks.

Beyond its importance in extreme-value theory, the same survival law emerges from a simple multiplicative mechanism. Specifically, if $X=V/W$, where $V$ and $W$ are independent random variables with distributions $\mathrm{Exponential}(\beta)$ (rate $\beta>0$) and $\mathrm{Gamma}(\alpha,1)$ (shape $\alpha>0$, unit rate), respectively, then conditioning on $W$ gives $\SF_X(x)=\e[e^{-\beta xW}]=(1+\beta x)^{-\alpha}$, $x\ge0$. This stochastic representation admits two complementary readings. In the language of survival analysis, $X$ is conditionally exponential given $W$ with hazard rate $\beta W$, so that $W$ acts as a gamma frailty \citep{Oakes1989}; gamma mixing of this type is likewise the standard device for describing unobserved heterogeneity in insurance claim data \citep{Boucher2008}. In the language of risk management, $X=V\cdot W^{-1}$ is the product of a light-tailed risk factor $V$ and an exogenous multiplicative factor $W^{-1}$, which places the Pareto law within the framework of multiplicative background risk models (MBRMs) \citep{Franke2006,Tsanakas2008,Asimit2016,Furman2020a}. From this viewpoint, heavy tails are not postulated but generated by the interaction between a light-tailed risk factor and a random multiplicative factor. It is this second reading that motivates the present paper.

An important advantage of this multiplicative representation is that it extends naturally to a multivariate framework, in which dependence can be introduced through background risk factors shared by several risks. Such extensions are particularly relevant in insurance, where risks are often interconnected and the joint occurrence of extreme risks drives pricing, reserving, reinsurance design, and solvency assessment. Multivariate models that combine Pareto margins with flexible dependence structures are of considerable actuarial interest. Numerous multivariate Pareto distributions have been proposed in the literature; see, for example, \cite{Chiragiev2009}, \cite{Asimit2010}, \cite{Sarabia2016}, \cite{Su2017b}, \cite{Su2017a}, and Chapter~6 of \cite{Arnold2015} for a comprehensive overview.

To make the discussion of MBRMs and their induced dependence structures more precise, let $N=\{1,\dots,n\}$, $n\in\mathbb{N}$, be a label set and let $\P(N)=\{I\subseteq N:I\neq\varnothing\}$ denote the set of non-empty subsets of $N$. Consider two collections $\A,\B\subseteq\P(N)$ and define $\A_i=\{I\in\A:i\in I\}$ and $\B_i=\{I\in\B:i\in I\}$ for each $i\in N$; throughout, $\A_i$ and $\B_i$ are assumed non-empty, so that all components introduced below are almost surely positive and finite. Let $\XX=(X_1,\dots,X_n)$ be a risk vector whose risk components admit the representation
\begin{equation}
\label{eq:main_rep}
X_i=\frac{V_i}{W_i},\qquad i\in N,
\end{equation}
where the numerator risk factor vector $\VV=(V_1,\dots,V_n)$ and the denominator risk factor vector $\WW=(W_1,\dots,W_n)$ are independent. Both may incorporate background risk factors that are idiosyncratic, affecting one risk component, or systemic, affecting several risk components. Dependence within either vector is permitted.

One approach introduces dependence among the risk components $(X_1,\ldots,X_n)$ through the denominator risk factor vector $\WW$. Suppose its coordinates are given by the additive gamma model
\begin{equation}
\label{eq:WW_Add_rep}
W_i=\sum_{I\in\A_i}G_I,\qquad i\in N,
\end{equation}
where the background risk factors $G_I$, $I\in\A$, are independent gamma variates with shape parameters $\alpha_I>0$ and unit rates. Each factor $G_I$ affects precisely the risk components indexed by $I$ and is idiosyncratic if it affects one component ($|I|=1$) and systemic if it affects several components ($|I|>1$). This specification is in the spirit of the additive multivariate gamma constructions of \cite{Mathai1991} and \cite{Furman2008}. If, in addition, $\VV$ consists of independent exponential risk factors with rates $\beta_i>0$, then conditioning on $\WW$ yields
\begin{equation}
\label{eq:6.2.8arnold}
\SF_{\XX}(x_1,\dots,x_n)
=
\e\left[\exp\left\{-\sum_{i\in N}\beta_i x_iW_i\right\}\right]
=
\prod_{I\in\A}
\left(1+\sum_{i\in I}\beta_i x_i\right)^{-\alpha_I},
\end{equation}
which is the multivariate distribution appearing in Equation~6.2.8 of \cite{Arnold2015}.

An alternative approach introduces dependence among the risk components $(X_1,\ldots,X_n)$ through the numerator risk factor vector $\VV$. Suppose that $\VV$ follows a Marshall--Olkin exponential distribution \citep{MarshallA.1967},
\begin{equation}
\label{eq:VV_MO_rep}
V_i=\min_{I\in\B_i}E_I,
\end{equation}
where the background risk factors $E_I$, $I\in\B$, are independent exponentials with rates $\beta_I>0$, each affecting the risk components indexed by $I$ through the minimum in \eqref{eq:VV_MO_rep}. As in the gamma risk factor system, $E_I$ is idiosyncratic if $|I|=1$ and systemic if $|I|>1$. If, furthermore, $\WW$ is comonotonic with common coordinate $W_i=W\sim\mathrm{Gamma}(\alpha,1)$, then
\begin{equation}
\label{eq:6.2.3arnold}
\SF_{\XX}(x_1,\dots,x_n)
=
\e\left[\exp\left\{-W\sum_{I\in\B}\beta_I
\max_{i\in I}\left(x_i\right)\right\}\right]
=
\left(
1+\sum_{I\in\B}\beta_I\max_{i\in I}\left(x_i\right)
\right)^{-\alpha},
\end{equation}
which corresponds to the model discussed in Equation~6.2.3 of \cite{Arnold2015}; see also \cite{Bernhart2013} for related scale mixtures of Marshall--Olkin laws in portfolio credit risk.

Although both constructions arise from the same multiplicative framework, they induce fundamentally different forms of dependence. The model in \eqref{eq:6.2.8arnold} generates dependence through the additive gamma structure in \eqref{eq:WW_Add_rep}: two risk components are dependent precisely when they share a gamma background risk factor. The resulting distribution is analytically tractable and flexible in modelling heterogeneous interactions among subsets of risk components, including independence. Nevertheless, it is absolutely continuous and therefore cannot accommodate singular components, such as exact ties, that is, equality of two or more risk components with positive probability. In contrast, the model in \eqref{eq:6.2.3arnold} incorporates additional dependence through systemic exponential background risk factors in \eqref{eq:VV_MO_rep}. The Marshall--Olkin mechanism is well established in reliability, default risk, and life insurance and naturally accommodates strong positive dependence as well as the singular behaviour associated with exact ties among risk components. However, because all risk components share the same gamma denominator $W$, they remain dependent even when $\B=\{\{1\},\ldots,\{n\}\}$, in which case \eqref{eq:6.2.3arnold} reduces to the classical multivariate Pareto distribution of the second kind \citep{Arnold2015} and the Pearson correlation between any two components equals $1/\alpha$ whenever $\alpha>2$. The strength of dependence is thus tied to the tail index: weak dependence is available only at the expense of lighter tails, and independence cannot be attained.

\subsection{A motivating example}
\label{subsec:motivating}
The general MBRM formulation above permits systemic background risk factors in either the numerator or the denominator. It therefore provides no structural reason to restrict attention to a common denominator, $W_i=W$, without also considering a common numerator, $V_i=V$. Whereas the former configuration underlies a classical multivariate Pareto construction, the latter, combined with component-specific gamma denominators, is not covered by either one of \eqref{eq:6.2.8arnold} and \eqref{eq:6.2.3arnold}, and, to the best of our knowledge, has not been studied in the actuarial literature.

A shared numerator with distinct denominators also has a natural behavioural interpretation. In delay \citep{Mazur1987} and probability \citep{Rachlin1991} discounting, the subjective value of a delayed or uncertain reward is commonly described by hyperbolic discount functions $(1+\beta x)^{-1}$ and their power generalizations $(1+\beta x)^{-\alpha}$, where $x$ is the delay or the odds against receipt. These are precisely Pareto survival functions, and the representation $X=V/W$ above shows that they arise from exponential discounting with a gamma-distributed rate. For rewards that are both delayed and probabilistic, \citet{Vanderveldt2015} find that the two discount functions combine multiplicatively, which corresponds to independent risk components in our framework, $\A=\B=\{\{1\},\{2\}\}$. The Gamma--Pareto sub-mixture nests this multiplicative benchmark and departs from it through an exponential background risk factor shared by the two dimensions, while dimension-specific gamma factors reflect the different mechanisms associated with delay and uncertainty. The simplest such departure, with a single shared exponential factor, already reveals the sub-mixture structure developed in the paper.

Let $n\ge2$, $\B=\{N\}$ and $\A=\{\{1\},\ldots,\{n\}\}$. Writing $V:=E_N\sim\mathrm{Exponential}(\beta_N)$, $W_i:=G_{\{i\}}$ and $\alpha_i:=\alpha_{\{i\}}$, the representation \eqref{eq:main_rep} becomes
\begin{equation}
\label{eq:motiv_model}
X_i=\frac{V}{W_i},\qquad
W_i\sim\mathrm{Gamma}(\alpha_i,1),\qquad i\in N,
\end{equation}
with $V,W_1,\ldots,W_n$ mutually independent. A single systemic exponential background risk factor $V$ is thus shared by $n$ risk components, each scaled by an idiosyncratic gamma background risk factor through the multiplier $W_i^{-1}$, $i\in N$. Every margin is Pareto, $\SF_{X_i}(x_i)=(1+\beta_Nx_i)^{-\alpha_i}$, yet the model is covered neither by \eqref{eq:6.2.8arnold}, which requires independent numerators, nor by \eqref{eq:6.2.3arnold}, which requires a common denominator. For $\xx=(x_1,\ldots,x_n)\in(0,\infty)^n$, conditioning on $\WW$ yields
\begin{equation}
\label{eq:joint-survival}
\SF_{\XX}(\xx)
=
\e\left[
\PP\left(V>x_1W_1,\ldots,V>x_nW_n\,\middle|\,\WW\right)
\right]
=
\e\left[
\exp\left\{-\beta_N\max_{i\in N}\left(x_iW_i\right)\right\}
\right].
\end{equation}
The product $x_iW_i$ is the level that the systemic risk factor $V$ must exceed for the $i$-th risk component to exceed $x_i$, so that joint exceedance occurs if and only if $V$ exceeds the largest of these levels. Accordingly, let $D_i(\xx)=\{x_iW_i\ge x_jW_j\ \text{for all}\ j\neq i\}$ denote the event that the exceedance constraint of the $i$-th risk component is \emph{binding}, $i\in N$. Since $\WW$ is absolutely continuous, ties occur with probability zero; hence the events $D_1(\xx),\ldots,D_n(\xx)$ partition the sample space up to a null set, so that $\sum_{i\in N}\PP(D_i(\xx))=1$. Decomposing \eqref{eq:joint-survival} along this partition gives the mixture
\begin{equation}
\label{eq:m}
\SF_{\XX}(\xx)
=
\sum_{i\in N}
\e\left[e^{-\beta_Nx_iW_i}\,\mathds{1}_{D_i(\xx)}\right]
=
\sum_{i\in N}
\PP\left(D_i(\xx)\right)\,
\e\left[e^{-\beta_Nx_iW_i}\,\middle|\,D_i(\xx)\right],
\end{equation}
whose components, however, are conditional Laplace transforms without a direct distributional meaning.

The key step is an exponential tilting of the gamma background risk factors. Because $\PP(X_i>x_i\mid\WW)=e^{-\beta_Nx_iW_i}$ and $D_i(\xx)$ is $\WW$-measurable, the $i$-th summand in \eqref{eq:m} equals $\PP(X_i>x_i,\,D_i(\xx))=\SF_{X_i}(x_i)\,\PP(D_i(\xx)\mid X_i>x_i)$. Moreover, conditioning on $\{X_i>x_i\}$ multiplies the joint density of $\WW$ by $e^{-\beta_Nx_iw_i}$, up to normalization, so that $W_i$ becomes $\mathrm{Gamma}(\alpha_i,1+\beta_Nx_i)$ distributed, whereas $W_j$, $j\neq i$, retain their laws and independence. Denoting by $W_i^\ast$ a random variable with the former law, independent of $W_j$, $j\neq i$, we arrive at
\begin{equation}
\label{eq:sub-m}
\SF_{\XX}(\xx)
=
\sum_{i\in N}p_i(\xx)\,\SF_{X_i}(x_i),
\qquad
p_i(\xx)
=
\PP\left(D_i(\xx)\,\middle|\,X_i>x_i\right)
=
\PP\left(D_i^\ast(\xx)\right),
\end{equation}
where $D_i^\ast(\xx)=\{x_iW_i^\ast\ge x_jW_j\ \text{for all}\ j\neq i\}$. The joint survival function of $\XX$ is thus a weighted sum of its Pareto marginal survival functions, the $i$-th weight being the probability that the $i$-th risk component's exceedance constraint is binding given that the component exceeds its own threshold.

Crucially, and in contrast to \eqref{eq:m}, the weights in \eqref{eq:sub-m} do not sum to one. By the scaling property of the gamma distribution we may take $W_i^\ast=W_i/(1+\beta_Nx_i)<W_i$, whence $D_i^\ast(\xx)\subseteq D_i(\xx)$. The difference $D_i(\xx)\setminus D_i^\ast(\xx)=\{x_iW_i/(1+\beta_Nx_i)<\max_{j\neq i}(x_jW_j)\le x_iW_i\}$ has positive probability because $\WW$ has a strictly positive density on $(0,\infty)^n$. Consequently, $p_i(\xx)<\PP(D_i(\xx))$ for every $i\in N$, so that $\sum_{i\in N}p_i(\xx)<1$, and \eqref{eq:sub-m} is a genuine \emph{sub-mixture} rather than a mixture. The deficit has a transparent probabilistic meaning: since $p_i(\xx)-\PP(D_i(\xx))=\Cov\big(\mathds{1}_{D_i(\xx)},\mathds{1}_{\{X_i>x_i\}}\big)/\SF_{X_i}(x_i)$,
\[
1-\sum_{i\in N}p_i(\xx)
=
-\sum_{i\in N}
\frac{
\Cov\left(
\mathds{1}_{D_i(\xx)},
\mathds{1}_{\{X_i>x_i\}}
\right)
}{
\SF_{X_i}(x_i)
}
>0.
\]
An exceedance of the $i$-th risk component signals a small value of $W_i$, hence a low level $x_iW_i$, which in turn makes it less likely that the $i$-th risk component's exceedance constraint is binding; the deficit aggregates precisely this negative association between marginal exceedance and dominance.

The weights also admit an appealing allocation interpretation. Let $S=\sum_{k\in N}W_k$ and $R_k=W_k/S$, $k\in N$. Then $\RR=(R_1,\ldots,R_n)\sim\mathrm{Dirichlet}(\alpha_1,\ldots,\alpha_n)$ is independent of $S$ \citep{Lukacs1955,Ng2011a}, and since the inequalities defining $D_i^\ast(\xx)$ are invariant under division by $S>0$,
\begin{equation}
\label{eq:Dirichlet_weights}
p_i(\xx)
=
\PP\left(
R_i\ge\left(1+\beta_Nx_i\right)\frac{x_j}{x_i}\,R_j
\ \text{for all}\ j\neq i
\right),
\qquad i\in N,
\end{equation}
while $\PP(D_i(\xx))$ is given by the same expression with the factor $1+\beta_Nx_i$ removed. Interpreting $R_i$ as the proportional contribution of the $i$-th idiosyncratic gamma background risk factor to the total gamma factor mass $S$, the weight $p_i(\xx)$ is the probability that its threshold-weighted contribution $x_iR_i$ exceeds every competing contribution $x_jR_j$ by at least the multiplicative factor $1+\beta_Nx_i$. This factor is the rate adjustment induced by conditioning on an exceedance of the $i$-th risk component. For instance, when $n=2$ and $\alpha_1=\alpha_2=1$, \eqref{eq:Dirichlet_weights} gives $p_i(x_1,x_2)=x_i/(x_1+x_2+\beta_Nx_1x_2)$, $i=1,2$, with deficit $\beta_Nx_1x_2/(x_1+x_2+\beta_Nx_1x_2)>0$, and one readily verifies that the resulting survival function belongs to neither of the families \eqref{eq:6.2.8arnold} and \eqref{eq:6.2.3arnold}.

\subsection{Contributions and outline}
\label{subsec:contributions}

The example in Section \ref{subsec:motivating} exhibits, in its simplest form, the phenomenon studied in this paper. We consider the representation in \eqref{eq:main_rep} in which the denominator risk factor vector $\WW$ follows the additive gamma structure in \eqref{eq:WW_Add_rep}, the numerator risk factor vector $\VV$ follows the Marshall--Olkin exponential structure in \eqref{eq:VV_MO_rep}, and $\VV$ and $\WW$ are independent. This formulation describes an MBRM with Pareto margins in which both risk factor systems may incorporate idiosyncratic and systemic background risk factors, extending the multiplicative frameworks studied in \cite{Tsanakas2008}, \cite{Asimit2016} and \cite{Furman2020a}. It contains \eqref{eq:6.2.8arnold}, \eqref{eq:6.2.3arnold} and \eqref{eq:motiv_model} as special cases. We call the resulting family the \emph{Gamma--Pareto sub-mixture}: its weights are generated by the gamma background risk factors, equivalently as Dirichlet probabilities, and its components are Pareto survival functions.

Our main result shows that, under an admissibility condition on the risk factor systems, the structure uncovered in \eqref{eq:sub-m} extends to this broader framework: the joint survival function of $\XX$ is a finite sub-mixture of Pareto survival functions, each combining the product structure of \eqref{eq:6.2.8arnold} with the maximum structure of \eqref{eq:6.2.3arnold}.
 These functions are indexed by orderings of classes of risk components, each class having a common gamma denominator, and are weighted by gamma, or equivalently Dirichlet, probabilities of the corresponding ordering regions after exponential tilting. Some Pareto components are effectively lower-dimensional, as in the motivating example. The contribution of each systemic exponential background risk factor affecting several classes is assigned to the class that imposes the largest effective threshold among them. This generalizes the motivating example, in which the shared numerator factor $V$ is assigned to the risk component imposing the largest effective threshold. When $\B=\{N\}$ and $\A=\{\{1\},\ldots,\{n\}\}$, grouping the ordering regions according to that component recovers the weights $p_i(\xx)$ and the representation in \eqref{eq:sub-m}.

We regard the principal novelty of the paper as a new interpretation of MBRMs with Pareto-distributed risk components. The conventional representation conditions on the gamma background risk factors: the individual risk components then have exponential conditional margins, and the joint survival function is obtained by averaging over these factors, as in \eqref{eq:6.2.8arnold}, \eqref{eq:6.2.3arnold} and \eqref{eq:joint-survival}. The sub-mixture representation offers a complementary reading. It expresses the joint survival function directly through Pareto survival functions and their Dirichlet weights. The weights are probabilities of orderings induced by the background risk factors and the risk thresholds, with exponential tilting accounting for the conditioning on exceedances of the corresponding auxiliary risk vectors. They also admit an allocation interpretation in terms of the proportional contributions of the gamma background risk factors to their total. In the motivating example, each weight is the probability that a particular risk component's exceedance constraint is binding, conditional on its marginal exceedance, and the sub-mixture deficit quantifies the association between these two events. In the general construction, the analogous interpretation involves the class orderings and exceedances of the auxiliary risk vectors. Since the weights depend on the arguments of the survival function and sum to at most one, the representation is a pointwise sub-mixture rather than a mixture of distributions with fixed weights. To the best of our knowledge, neither this Dirichlet sub-mixture representation nor its allocation interpretation has previously appeared in the literature on multivariate Pareto distributions or background risk models.

Beyond this interpretation, the proposed Gamma--Pareto sub-mixture enlarges the class of tractable models for dependent risk components with Pareto margins and encompasses many known multivariate Pareto constructions. We show that it is closed under marginalization, possesses Pareto margins and tractable distributions of minima, and admits a complete characterization of independence and, under a non-degeneracy condition on the gamma risk factor system, of its singular and absolutely continuous parts, so that the two may coexist within a single model. We further prove that the model is associated, decompose its covariances into the contributions of the Marshall--Olkin and additive gamma risk factor structures and their interaction, and describe its extremal behaviour through regular variation along rays, tail dependence coefficients and the coefficient of tail dependence, all of which bear directly on quantitative risk management. The extremal analysis reveals a further distinction: the gamma risk factors determine the presence of asymptotic tail dependence and the decay rates of joint exceedance probabilities, while the exponential risk factors affect the associated tail constants.

The remainder of the paper is organized as follows. Section~\ref{sec:GammaPareto} introduces the proposed model, establishes the sub-mixture representation, and derives its principal distributional properties. Section~\ref{sec:Dependence_Measures} studies its dependence structure and the associated dependence measures. Finally, Section~\ref{sec:conclusions} concludes the paper. The more technical proofs are provided in Appendix~\ref{app:proofs_sec2}, and Appendix~\ref{app:examples} works through two examples that illustrate the main results.
\section{The Gamma--Pareto sub-mixture}
\label{sec:GammaPareto}

The motivating example of Section~\ref{subsec:motivating} contains, in embryonic form, all ingredients of the general construction: a partition of the latent space according to which risk component's exceedance constraint is binding, an exponential tilting of the gamma background risk factors, and a sub-mixture of Pareto survival functions. The purpose of this section is to show that these ingredients extend to the general framework under an admissibility condition on the risk factor systems. We first formalize the interplay between the factor systems $\A$ and $\B$ in Section~\ref{subsec:framework}, then state and interpret the sub-mixture representation in Section~\ref{subsec:representation}, and finally derive the principal structural properties of the resulting family in Sections~\ref{subsec:structural}--\ref{subsec:examples}. Throughout, $\xx=(x_1,\dots,x_n)$ denotes a point of $[0,\infty)^n$, we retain the standing assumption that $\A_i$ and $\B_i$ are non-empty for every $i\in N$, and the longer proofs are collected in Appendix~\ref{app:proofs_sec2}.

\subsection{Factor-selection framework}
\label{subsec:framework}

As defined in \eqref{eq:WW_Add_rep} and \eqref{eq:VV_MO_rep}, the collections $\A$ and $\B$ determine the factor selections governing the denominator risk factor vector $\WW$ and the numerator risk factor vector $\VV$, respectively; for each $i\in N$, the sets $\A_i$ and $\B_i$ identify the factors driving $W_i$ and $V_i$. For brevity, we refer to the background risk factors $G_I$, $I\in\A$, and $E_J$, $J\in\B$, as gamma and exponential risk factors, to $I$ and $J$ as their labels, and to $\A$ and $\B$ as the (risk) factor systems. From the perspective of the denominator, two indices are indistinguishable when they are affected by exactly the same gamma factors. Formally, let $M=\{1,\dots,m\}$, $m\le n$, label the equivalence classes $N_1,\ldots,N_m$ of $N$ induced by the relation
\[
i\sim j \qquad\Longleftrightarrow\qquad \A_i=\A_j.
\]
The classes form a partition of $N$, and the class-level collection $\AA_k:=\A_i$, $i\in N_k$, is well defined. For $i\in N$, we denote by $k(i)\in M$ the label of the class containing $i$. Consequently, all components within a class share a common denominator,
\begin{equation}
\label{eq:class_denominator}
W_i=W^{(k)}:=\sum_{I\in\AA_k}G_I,\qquad i\in N_k,\quad k\in M.
\end{equation}
For any $I\in\A\cup\B$, we define its class support $M_I=\{k\in M:I\cap N_k\neq\varnothing\}$ as the set of classes affected by the risk factor indexed by $I$. Every $I\in\A$ is a union of classes: if $i\in I$ and $j\sim i$, then $I\in\A_i=\A_j$, whence $j\in I$. Therefore, for all $I\in\A$ and $k\in M$,
\begin{equation}
\label{eq:incidence}
I\in\AA_k
\quad\Longleftrightarrow\quad
k\in M_I
\quad\Longleftrightarrow\quad
N_k\subseteq I,
\end{equation}
an identity that will be used repeatedly.

The joint survival event can be written as $\{V_i>x_iW_i\ \forall i\in N\}=\{E_J>\max_{j\in J}x_jW_j\ \forall J\in\B\}$, and by \eqref{eq:class_denominator},
\[
\max_{j\in J}\left(x_jW_j\right)=\max_{k\in M_J}\Bigl(\max_{j\in J\cap N_k}\left(x_j\right)W^{(k)}\Bigr),\qquad J\in\B.
\]
If an exponential risk factor with support $J$ affects several classes but only part of some class, the threshold attached to that class depends on $J$; different exponential factors may then rank the classes differently, and no single ordering of the classes resolves all maxima simultaneously. To ensure a coherent interaction between the numerator and the denominator, we therefore require every element of $\B$ to be \emph{admissible}: a subset $I\subseteq N$ is admissible if either $I\subseteq N_k$ for some $k\in M$ or $I=\bigcup_{k\in M_I}N_k$. Accordingly, $\B$ admits the disjoint decomposition $\B=\B_{-}\cup\B_{+}$, where
\[
\B_{-}=\left\{I\in\B:I\subseteq N_k\text{ for some }k\in M\right\}
\quad\text{and}\quad
\B_{+}=\left\{I\in\B:I=\bigcup_{k\in M_I}N_k,\ |M_I|\ge2\right\}.
\]
Thus $\B_{-}$ collects the exponential risk factors confined to a single class, possibly the whole class, and for $J\in\B_{-}$ we denote by $k_J$ the unique element of $M_J$. The collection $\B_{+}$ collects the exponential risk factors acting on at least two entire classes, for which the maximum above involves only the \emph{class thresholds}
\begin{equation}
\label{eq:class_thresholds}
x_k^{*}:=\max_{j\in N_k}\left(x_j\right),\qquad k\in M.
\end{equation}
This is precisely what renders a single ordering of the classes sufficient. No such restriction is needed for $\A$, whose elements are automatically unions of classes by \eqref{eq:incidence}.

The multivariate Pareto distributions discussed in Section~\ref{sec:introd} arise as special cases of this framework. The distribution in \eqref{eq:6.2.8arnold} is recovered by taking an arbitrary collection $\A\subseteq\P(N)$ together with $\B=\{\{1\},\ldots,\{n\}\}$, whereas the distribution in \eqref{eq:6.2.3arnold} corresponds to $\A=\{N\}$, so that $m=1$, and $\B\subseteq\P(N)$ arbitrary; in both cases $\B_{-}=\B$ and $\B_{+}=\varnothing$. The motivating example \eqref{eq:motiv_model}, in contrast, corresponds to $\A=\{\{1\},\ldots,\{n\}\}$ and $\B=\{N\}$, so that $m=n$, $N_k=\{k\}$, $\B_{-}=\varnothing$ and $\B_{+}=\{N\}$. The classical constructions thus involve only exponential risk factors confined to a single class; the collection $\B_{+}$ introduces systemic exponential background risk factors affecting classes with different gamma denominators.

\subsection{The sub-mixture representation}
\label{subsec:representation}

For $\xx\in[0,\infty)^n$, let $M_+=M_+(\xx)=\{k\in M:x_k^{*}>0\}$ be the set of active classes, $m_+=|M_+|$, and $\B_{+}(\xx)=\{J\in\B_{+}:M_J\cap M_+\neq\varnothing\}$ the set of labels of active exponential risk factors affecting multiple classes. Denote by $\SS_{M_+}$ the set of bijective enumerations $\pi:\{1,\dots,m_+\}\to M_+$, and, for $\pi\in\SS_{M_+}$ and $J\in\B_{+}(\xx)$, let
\[
k_J^{\pi}=\pi\Bigl(\min\bigl\{t\in\{1,\dots,m_+\}:\pi(t)\in M_J\cap M_+\bigr\}\Bigr)
\]
be the \emph{leading class} of $J$, that is, the first active class affected by the exponential risk factor indexed by $J$ under the ordering $\pi$. For $\xx=\mathbf{0}$, $\SS_{\varnothing}$ consists of the empty enumeration only, and \eqref{eq:GammaPareto} below reduces to $\SF_{\XX}(\mathbf{0})=1$.

\begin{theorem}
\label{thm:mainPD}
The joint survival function $\SF_{\XX}$ of $\XX$ admits a finite sub-mixture representation in terms of Pareto survival functions, each weighted by a gamma, or equivalently Dirichlet, probability; see \eqref{eq:Dirichlet_general}. Formally, for every $\xx\in[0,\infty)^n$,
\begin{equation}
\label{eq:GammaPareto}
\SF_{\XX}(\xx)=\sum_{\pi\in\SS_{M_+}}\PP\left(\GG\in\Omega^{\pi}(\xx)\right)\prod_{I\in\A}\left(1+\Lambda_I^{\pi}(\xx)\right)^{-\alpha_I},
\end{equation}
with
\begin{equation}
\label{eq:GammaPareto_weights_lessthanone}
\sum_{\pi\in\SS_{M_+}}\PP\left(\GG\in\Omega^{\pi}(\xx)\right)\le 1,
\end{equation}
where $\GG=(G_I,I\in\A)$ is the vector of independent gamma variates with shapes $\alpha_I$ and unit rates,
\begin{equation}
\label{eq:Lambda_MP}
\Lambda_I^{\pi}(\xx)=\sum_{\substack{J\in\B_{-}\\k_J\in M_I}}\beta_J\max_{j\in J}\left(x_j\right)+\sum_{\substack{J\in\B_{+}(\xx)\\k_J^{\pi}\in M_I}}\beta_J\,x^{*}_{k_J^{\pi}},\qquad I\in\A,
\end{equation}
and $\Omega^{\pi}(\xx)\subseteq[0,\infty)^{|\A|}$ is the polyhedral cone of points $\gg=(g_I,I\in\A)$ satisfying
\begin{equation}
\label{eq:Omega}
x^{*}_{\pi(1)}\sum_{I\in\AA_{\pi(1)}}\frac{g_I}{1+\Lambda_I^{\pi}(\xx)}\ge\cdots\ge x^{*}_{\pi(m_+)}\sum_{I\in\AA_{\pi(m_+)}}\frac{g_I}{1+\Lambda_I^{\pi}(\xx)}.
\end{equation}
\end{theorem}

The proof, given in Appendix~\ref{app:proofs_sec2}, follows the three steps of the motivating example. First, conditioning on the factors yields $\SF_{\XX}(\xx)=\e[\exp\{-\sum_{J\in\B}\beta_J\max_{j\in J}(x_jW_j)\}]$. Second, the latent space is partitioned according to the ordering of the effective class thresholds $x_k^{*}W^{(k)}$, that is, into the cones
\begin{equation}
\label{eq:Omega_tilde}
\widetilde{\Omega}^{\pi}(\xx)=\left\{\gg\in[0,\infty)^{|\A|}:\ x^{*}_{\pi(1)}\sum_{I\in\AA_{\pi(1)}}g_I\ge\cdots\ge x^{*}_{\pi(m_+)}\sum_{I\in\AA_{\pi(m_+)}}g_I\right\},\qquad\pi\in\SS_{M_+},
\end{equation}
which cover $[0,\infty)^{|\A|}$ and overlap only on a Lebesgue-null set. On $\{\GG\in\widetilde{\Omega}^{\pi}(\xx)\}$, the maximum attached to each $J\in\B_{+}(\xx)$ is attained at its leading class $k_J^{\pi}$. The contribution of the exponential background risk factor is thus assigned to the class imposing the largest effective threshold among those it affects, and by \eqref{eq:incidence} the exponent becomes the linear form $\sum_{I\in\A}\Lambda_I^{\pi}(\xx)G_I$. Third, an exponential tilting of $\GG$ converts $e^{-\sum_I\Lambda_I^{\pi}G_I}$ into the product of Pareto survival terms in \eqref{eq:GammaPareto} and maps $\widetilde{\Omega}^{\pi}(\xx)$ onto $\Omega^{\pi}(\xx)=\{\gg:(g_I/(1+\Lambda_I^{\pi}(\xx)))_{I\in\A}\in\widetilde{\Omega}^{\pi}(\xx)\}$. The inequality \eqref{eq:GammaPareto_weights_lessthanone} is more delicate: because $\Lambda^{\pi}$ depends on $\pi$, the cones $\Omega^{\pi}(\xx)$ are rescaled differently for different orderings, and it must be shown that they nevertheless have pairwise disjoint interiors.

\begin{remark}
\label{rm:GammaPareto_interpretation}
The weights and components in \eqref{eq:GammaPareto} admit a probabilistic interpretation that extends the one in Section~\ref{subsec:motivating}. Let $\xx\in(0,\infty)^n$, so that $M_+=M$, and fix $\pi\in\SS_M$. Assigning the contribution of each exponential risk factor indexed by $J\in\B_{+}$ to its leading class yields the factor system $\B^{\pi}=\B_{-}\cup\{N_{k_J^{\pi}}:J\in\B_{+}\}$, in which every exponential risk factor is confined to a single class; factors acting on the same set are merged by adding their rates. Let the auxiliary risk vector $\XX^{\pi}=(V_i^{\pi}/W_i,\,i\in N)$ be built from the same $\WW$ and an independent Marshall--Olkin vector $\VV^{\pi}$ with factor system $\B^{\pi}$, with the convention $V_i^{\pi}=+\infty$ if no exponential risk factor in the system $\B^{\pi}$ affects $i$. Since every exponential risk factor in the system $\B^{\pi}$ is confined to a single class, conditioning on $\GG$ gives $\PP(\XX^{\pi}>\xx\mid\GG)=\exp\{-\sum_{I\in\A}\Lambda_I^{\pi}(\xx)G_I\}$. Two consequences follow. First, the $\pi$-th component in \eqref{eq:GammaPareto} equals $\SF_{\XX^{\pi}}(\xx)$, the survival function of a Pareto vector of the hybrid form \eqref{eq:BplusEmpty_GammaParetoHybrid} below; coordinates not affected by any exponential risk factor in $\B^{\pi}$ are degenerate at $+\infty$, so that some components are effectively lower-dimensional. Second, conditionally on $\{\XX^{\pi}>\xx\}$, the variables $G_I$ are independent and $\mathrm{Gamma}(\alpha_I,1+\Lambda_I^{\pi}(\xx))$ distributed, whence
\begin{equation}
\label{eq:weights_conditional}
\PP\left(\GG\in\Omega^{\pi}(\xx)\right)=\PP\left(\GG\in\widetilde{\Omega}^{\pi}(\xx)\,\middle|\,\XX^{\pi}>\xx\right).
\end{equation}
Each weight is thus the probability that the latent ordering of the effective class thresholds is $\pi$, given that the auxiliary risk vector $\XX^{\pi}$ exceeds $\xx$. As $\sum_{\pi}\PP(\GG\in\widetilde{\Omega}^{\pi}(\xx))=1$, the sub-mixture deficit admits the covariance representation
\[
1-\sum_{\pi\in\SS_M}\PP\left(\GG\in\Omega^{\pi}(\xx)\right)
=-\sum_{\pi\in\SS_M}\frac{\Cov\left(\mathds{1}_{\{\GG\in\widetilde{\Omega}^{\pi}(\xx)\}},\mathds{1}_{\{\XX^{\pi}>\xx\}}\right)}{\SF_{\XX^{\pi}}(\xx)},
\]
which quantifies the dependence between the latent orderings and the exceedances of the auxiliary risk vectors. In the motivating example, $\XX^{\pi}$ reduces to the single coordinate $X_{\pi(1)}$, and the $(n-1)!$ orderings sharing the leader $i$ aggregate to $p_i(\xx)=\PP(D_i(\xx)\mid X_i>x_i)$, recovering \eqref{eq:sub-m}.
\end{remark}

\begin{remark}
\label{rm:GammaPareto_BplusEmpty}
For each $\xx$, representation \eqref{eq:GammaPareto} expresses $\SF_{\XX}(\xx)$ as a sub-convex combination of at most $m_+!$ multivariate Pareto survival functions. Since both the weights and the components depend on $\xx$, this is a pointwise sub-mixture rather than a mixture of distributions with fixed weights; it is precisely this feature that allows finitely many Pareto laws to reproduce a joint law that is itself not of Pareto product form. The deficit in \eqref{eq:GammaPareto_weights_lessthanone} equals the gamma probability of the region left uncovered by the cones $\Omega^{\pi}(\xx)$. Because these cones are rescaled differently for different $\pi$, gaps may open between them, which motivates the term \emph{sub-mixture}. The inequality is typically strict; in the motivating example it is strict for every $\xx\in(0,\infty)^n$. Equality may nevertheless hold even when $\B_{+}\neq\varnothing$, namely when the leading class of every exponential risk factor affecting multiple classes is determined almost surely. For instance, if $n=2$, $\A=\{\{1,2\},\{2\}\}$ and $\B=\{\{1,2\}\}$, then $W_1=G_{\{1,2\}}\le G_{\{1,2\}}+G_{\{2\}}=W_2$; for $x_1\le x_2$ the second class leads almost surely and the weights sum to one, whereas for $x_1>x_2$ their sum is strictly smaller than one.

If $\B_{+}=\varnothing$, then $\Lambda_I^{\pi}=\Lambda_I$ does not depend on $\pi$, and the cones $\Omega^{\pi}(\xx)$, $\pi\in\SS_{M_+}$, are the ordering cones of a single family of scores. They therefore partition $[0,\infty)^{|\A|}$ up to a null set, the weights sum to one, and \eqref{eq:GammaPareto} simplifies to
\begin{equation}
\label{eq:BplusEmpty_GammaParetoHybrid}
\SF_{\XX}(\xx)
=
\prod_{I\in\A}
\left(
1+\Lambda_I(\xx)
\right)^{-\alpha_I}
=
\prod_{I\in\A}
\Biggl(
1+
\sum_{\substack{J\in\B\\k_J\in M_I}}
\beta_J
\max_{j\in J}\left(x_j\right)
\Biggr)^{-\alpha_I}.
\end{equation}
This special case is a unified multivariate Pareto model encompassing the distributions in \eqref{eq:6.2.8arnold} and \eqref{eq:6.2.3arnold}. By Remark~\ref{rm:GammaPareto_interpretation}, every component of the general sub-mixture \eqref{eq:GammaPareto} is of this hybrid form, with $\B$ replaced by the auxiliary factor system $\B^{\pi}$.
\end{remark}

The sub-mixture weights also admit an equivalent Dirichlet representation. Let $S$ denote the aggregate $\sum_{I\in\A}G_I$ and set $R_I=G_I/S$ for $I\in\A$. The vector $\RR=(R_I,I\in\A)$ is supported on the standard simplex $\Delta^{|\A|-1}$, follows the Dirichlet distribution with parameters $(\alpha_I,I\in\A)$, and is independent of $S$ \citep{Ng2011a}. Since the cones $\Omega^{\pi}(\xx)$ are defined by inequalities that are homogeneous of degree one in $\gg$, membership is invariant under positive scaling, and therefore
\begin{equation}
\label{eq:Dirichlet_general}
\PP\left(\GG\in\Omega^{\pi}(\xx)\right)=\PP\left(\RR\in\Omega^{\pi}(\xx)\cap\Delta^{|\A|-1}\right),\qquad\pi\in\SS_{M_+}.
\end{equation}
Geometrically, each weight is the Dirichlet probability of a convex polytope in the simplex, and the deficit is the Dirichlet probability of the part of the simplex left uncovered by these polytopes. This is the general counterpart of the Dirichlet allocation interpretation in \eqref{eq:Dirichlet_weights}: $R_I$ measures the proportional contribution of the factor $G_I$ to the aggregate gamma factor mass $S$. Example~\ref{ex:B_regional} in Appendix~\ref{app:examples} works through these objects for a trivariate risk vector, in which the weights reduce to incomplete beta functions along the diagonal.

\begin{remark}
\label{rm:GammaPareto_Dirichlet}
The gamma and Dirichlet representations offer complementary routes to computing the sub-mixture weights. The gamma representation is generally preferable for Monte Carlo simulation. Its components are independent, so that sampling is straightforward, fully parallelizable and scalable to high dimensions, and a single gamma sample can be reused to evaluate all weights at all arguments $\xx$ through the explicit linear inequalities \eqref{eq:Omega}. Its principal limitation is the unbounded integration domain, which makes deterministic quadrature considerably more challenging. The Dirichlet representation, by contrast, is supported on the compact simplex, which is advantageous for deterministic numerical integration. This benefit is offset by the dependence among the Dirichlet components and by the geometric complexity of the polytopes $\Omega^{\pi}(\xx)\cap\Delta^{|\A|-1}$, whose evaluation requires polyhedral decomposition or triangulation. The Dirichlet route is therefore practical mainly in low dimensions; for $|\A|=2$ it yields closed forms in terms of incomplete beta functions, as illustrated in Example~\ref{ex:GammaPareto_Bivariate}.
\end{remark}

\subsection{Margins, minima and independence}
\label{subsec:structural}

The sub-mixture representation provides a transparent framework for studying the structural properties of the proposed family. We begin with its behaviour under marginalization, which also yields the univariate margins.

\begin{proposition}
\label{prop:GammaPareto_marginalization}
The Gamma--Pareto sub-mixture model is closed under marginalization. Specifically, for any non-empty $L\subseteq N$, let
\[
\A^{L}=\{I\cap L:I\in\A,\ I\cap L\neq\varnothing\},\qquad
\B^{L}=\{I\cap L:I\in\B,\ I\cap L\neq\varnothing\},
\]
and, for $I\in\A^{L}$ and $I'\in\B^{L}$,
\[
\alpha_I^{L}=\sum_{\substack{J\in\A\\J\cap L=I}}\alpha_J,\qquad
\beta_{I'}^{L}=\sum_{\substack{J\in\B\\J\cap L=I'}}\beta_J.
\]
Then $\XX^{L}=(X_i,\,i\in L)$ is a Gamma--Pareto sub-mixture on the label set $L$ with factor systems $(\A^{L},\B^{L})$, shapes $(\alpha_I^{L})$ and rates $(\beta_{I'}^{L})$. Its survival function is given by Theorem~\ref{thm:mainPD} applied to this system, and it coincides with $\SF_{\XX}(\xx)$ evaluated at $x_j=0$, $j\notin L$.
\end{proposition}

\begin{proof}
For $I\in\A^{L}$ and $I'\in\B^{L}$, define $G_I^{L}=\sum_{J\in\A:\,J\cap L=I}G_J$ and $E_{I'}^{L}=\min_{J\in\B:\,J\cap L=I'}E_J$. These families are independent, being built from disjoint groups of independent variables, with $G_I^{L}\sim\mathrm{Gamma}(\alpha_I^{L},1)$ and $E_{I'}^{L}\sim\mathrm{Exponential}(\beta_{I'}^{L})$. For $i\in L$, a set $J\in\A$ contains $i$ if and only if $J\cap L\in\A^{L}_i$, so that $W_i=\sum_{I\in\A^{L}_i}G_I^{L}$ and, similarly, $V_i=\min_{I'\in\B^{L}_i}E^{L}_{I'}$, where $\A_i^{L}$ and $\B_i^{L}$ are non-empty. It remains to verify the admissibility of $\B^{L}$. For $i,j\in L$, we have $\A^{L}_i=\A^{L}_j$ if and only if $\A_i=\A_j$, because any $J\in\A_i\setminus\A_j$ yields $J\cap L\in\A^{L}_i\setminus\A^{L}_j$. Hence the classes of $L$ are the non-empty sets $N_k\cap L$. If $J\in\B_{-}$, then $J\cap L\subseteq N_{k_J}\cap L$, whereas if $J\in\B_{+}$, then $J\cap L=\bigcup_{k\in M_J}(N_k\cap L)$ is a union of classes of $L$. Thus every element of $\B^{L}$ is admissible, and Theorem~\ref{thm:mainPD} applies. The last assertion holds because $\{X_j>0\}$ has probability one for every $j\in N$.
\end{proof}

The last assertion of Proposition~\ref{prop:GammaPareto_marginalization} explains why Theorem~\ref{thm:mainPD} is formulated on $[0,\infty)^n$ with the set $M_+(\xx)$ of active classes: margins are obtained simply by switching off the corresponding arguments. In particular, for $L=\{i\}$ we have $\A^{L}=\B^{L}=\{\{i\}\}$, a single class and a single ordering, and
\[
\SF_{X_i}(x_i)=\left(1+\sol{\beta}_i x_i\right)^{-\sol{\alpha}_i},
\qquad
\sol{\alpha}_i=\sum_{I\in\A_i}\alpha_I,
\quad
\sol{\beta}_i=\sum_{J\in\B_i}\beta_J.
\]
Each margin thus belongs to the Pareto type~II (Lomax) family with shape $\sol{\alpha}_i$ and rate $\sol{\beta}_i$, with
\[
\e[X_i]=\frac{1}{\sol{\beta}_i(\sol{\alpha}_i-1)},\quad\sol{\alpha}_i>1,
\qquad\text{and}\qquad
\Var[X_i]=\frac{\sol{\alpha}_i}{\sol{\beta}_i^{\,2}(\sol{\alpha}_i-1)^2(\sol{\alpha}_i-2)},\quad\sol{\alpha}_i>2.
\]
We also write $\sol{\alpha}^{(k)}=\sum_{I\in\AA_k}\alpha_I$ for the shape of the class denominator $W^{(k)}$, so that $\sol{\alpha}_i=\sol{\alpha}^{(k)}$ whenever $i\in N_k$, and, for $i\neq j$, $\sol{\alpha}_{ij}=\sum_{I\in\A_i\cap\A_j}\alpha_I$ and $\sol{\beta}_{ij}=\sum_{J\in\B_i\cap\B_j}\beta_J$ for the aggregate shape and rate of the factors shared by $X_i$ and $X_j$.

Minima are important in actuarial practice, for instance in first-to-default and joint-life products, and they are handled with equal ease. Since $\{\min_{i\in L}X_i>t\}=\{X_i>t\ \forall i\in L\}$, the survival function of a minimum is the diagonal section of the corresponding marginal survival function.

\begin{proposition}
\label{prop:GammaPareto_Minimum}
Let $L\subseteq N$ be non-empty, let $T^{L}:=\min_{i\in L}X_i$, and let $M^{L}=\{k\in M:N_k\cap L\neq\varnothing\}$. Then, for every $t>0$,
\begin{equation}
\label{eq:minSF}
\SF_{T^{L}}(t)
=\sum_{\pi\in\SS_{M^{L}}}\PP\left(\GG^{L}\in\Omega^{\pi}(t\one_L)\right)
\prod_{I\in\A^{L}}\left(1+t\,\Lambda_I^{\pi}(\one_L)\right)^{-\alpha^{L}_I},
\end{equation}
where $\one_L=(1,\dots,1)\in\R^{|L|}$, $\GG^{L}=(G^{L}_I,I\in\A^{L})$, and $\Omega^{\pi}$ and $\Lambda_I^{\pi}$ are those of Theorem~\ref{thm:mainPD} for the factor system $(\A^{L},\B^{L})$ of Proposition~\ref{prop:GammaPareto_marginalization}. In particular, if $L\subseteq N_k$ for some $k\in M$, then $T^{L}$ is Lomax distributed with
\[
\SF_{T^{L}}(t)=\bigl(1+\sol{\beta}^{L}\,t\bigr)^{-\sol{\alpha}^{(k)}},
\qquad
\sol{\beta}^{L}:=\sum_{\substack{J\in\B\\J\cap L\neq\varnothing}}\beta_J.
\]
\end{proposition}

\begin{proof}
Since $\SF_{T^{L}}(t)=\SF_{\XX^{L}}(t\one_L)$, the first assertion follows from Proposition~\ref{prop:GammaPareto_marginalization} and Theorem~\ref{thm:mainPD}. For $t>0$ every class of $L$ is active, and $\Lambda_I^{\pi}(t\one_L)=t\,\Lambda_I^{\pi}(\one_L)$ by the positive homogeneity of \eqref{eq:Lambda_MP}. If $L\subseteq N_k$, all components of $\XX^{L}$ share the denominator $W^{(k)}$, so that $T^{L}=\min_{J\in\B:\,J\cap L\neq\varnothing}E_J/W^{(k)}$. The numerator is exponential with rate $\sol{\beta}^{L}$ and independent of $W^{(k)}\sim\mathrm{Gamma}(\sol{\alpha}^{(k)},1)$, and the Lomax form follows by conditioning on $W^{(k)}$.
\end{proof}

At the opposite end of the dependence spectrum from comonotonicity, the model is also capable of generating independence among subsets of components. The following proposition shows that independence occurs precisely when no factor is shared.

\begin{proposition}
\label{prop:GammaPareto_Independence}
Let $L_1,L_2\subseteq N$ be disjoint. Then the sub-vectors
\[
\XX^{L_1}:=(X_i,\ i\in L_1),\qquad \XX^{L_2}:=(X_i,\ i\in L_2)
\]
are independent if and only if
\[
\left(\bigcup_{i\in L_1}\A_i\right)\bigcap\left(\bigcup_{i\in L_2}\A_i\right)=\varnothing\quad\text{and}\quad\left(\bigcup_{i\in L_1}\B_i\right)\bigcap\left(\bigcup_{i\in L_2}\B_i\right)=\varnothing.
\]
\end{proposition}

\begin{proof}
If both intersections are empty, the two sub-vectors are built from disjoint families of independent factors $G_I$ and $E_J$, and are therefore independent. Conversely, if one of the intersections is non-empty, there exist $i\in L_1$ and $j\in L_2$ with $\A_i\cap\A_j\neq\varnothing$ or $\B_i\cap\B_j\neq\varnothing$, i.e., with $\sol{\alpha}_{ij}>0$ or $\sol{\beta}_{ij}>0$. By Proposition~\ref{prop:GammaPareto_marginalization}, for $x_i,x_j>0$,
\begin{align*}
\PP(X_i>x_i,\,X_j>x_j)&=\e\left[\exp\left\{-\left(\sol{\beta}_{i}-\sol{\beta}_{ij}\right)x_iW_i-\left(\sol{\beta}_{j}-\sol{\beta}_{ij}\right)x_jW_j-\sol{\beta}_{ij}\max\left(x_iW_i,x_jW_j\right)\right\}\right]
\\
&\ge \e\left[\exp\left\{-\sol{\beta}_{i}\,x_iW_i-\sol{\beta}_{j}\,x_jW_j\right\}\right]
\\
&=\left(1+\sol{\beta}_{i}x_i\right)^{-\sol{\alpha}_{i}+\sol{\alpha}_{ij}}\left(1+\sol{\beta}_{j}x_j\right)^{-\sol{\alpha}_{j}+\sol{\alpha}_{ij}}\left(1+\sol{\beta}_{i}x_i+\sol{\beta}_{j}x_j\right)^{-\sol{\alpha}_{ij}}
\\
&\ge\left(1+\sol{\beta}_{i}x_i\right)^{-\sol{\alpha}_{i}}\left(1+\sol{\beta}_{j}x_j\right)^{-\sol{\alpha}_{j}}
=\PP(X_i>x_i)\,\PP(X_j>x_j),
\end{align*}
where the first inequality uses $\max(a,b)\le a+b$ and is strict if $\sol{\beta}_{ij}>0$, and the second uses $1+a+b\le(1+a)(1+b)$ for $a,b\ge0$ and is strict if $\sol{\alpha}_{ij}>0$. Hence $(X_i,X_j)$ is dependent, and so are $\XX^{L_1}$ and $\XX^{L_2}$.
\end{proof}

The proof shows, in addition, that every pair of components is positively quadrant dependent, a first manifestation of the positive dependence studied in Section~\ref{sec:Dependence_Measures}. Taking $L_1=\{i\}$ and $L_2=\{j\}$ for all $i\neq j$, and noting that disjointness of all factor sets implies mutual independence, the components $X_1,\ldots,X_n$ are mutually independent if and only if $\A_i\cap\A_j=\varnothing$ and $\B_i\cap\B_j=\varnothing$ for all $i\neq j\in N$. Since every $\A_i$ and $\B_i$ is non-empty, this is equivalent to $\A=\B=\{\{1\},\ldots,\{n\}\}$.

\subsection{Singular and absolutely continuous components}
\label{subsec:singular}

Beyond absolutely continuous dependence, the Gamma--Pareto sub-mixture may place positive mass on lower-dimensional sets. The prototypical mechanism arises from the interplay of the two factor systems. Within a class $N_k$, all components share the denominator $W^{(k)}$, so that $X_i=X_j$ if and only if $V_i=V_j$ for $i\neq j\in N_k$; ties within a class are thus inherited from $\VV$. If some exponential risk factor indexed by $J\in\B$ affects both $i$ and $j$, the Marshall--Olkin event $\{V_i=V_j\}$ has positive probability, since it occurs whenever $E_J$ arrives first among the exponential risk factors affecting $i$ or $j$, and it is transported to the diagonal $\{X_i=X_j\}$. Components in different classes, by contrast, tie with probability zero. For $i\in N_k$ and $j\in N_{\ell}$ with $k\neq\ell$, the event $\{X_i=X_j\}=\{V_iW^{(\ell)}=V_jW^{(k)}\}$ is, conditionally on $\VV$, a non-trivial linear constraint on the absolutely continuous vector $\GG$, because $\AA_k\neq\AA_\ell$. The following proposition quantifies the within-class singular mass.

\begin{proposition}
\label{prop:GammaPareto_singular}
For any class $N_k$, $k\in M$, let $L\subseteq N_k$ with $|L|\ge2$ be such that $\bigcap_{i\in L}\B_i\neq\varnothing$. Define the exact-tie event
\[
\Theta_k^L=\left\{X_i=X_j,\ \forall i,j\in L\right\}.
\]
Then
\[
\PP\left(\Theta_k^L\right)=\frac{\sum\limits_{\substack{J\in\B\\L\subseteq J}}\beta_J}{\sum\limits_{\substack{J\in\B\\ J\cap L\neq\varnothing}}\beta_J}>0,
\]
and, for any $i\in L$, the conditional survival function of $X_i$ given $\Theta_k^L$ is
\[
\SF_{X_i\mid\Theta_k^L}(x)=\left(1+\sol{\beta}^{L}\, x\right)^{-\sol{\alpha}^{(k)}},\qquad x\ge0,
\]
with $\sol{\alpha}^{(k)}=\sum_{I\in\AA_k}\alpha_I$ and $\sol{\beta}^{L}=\sum_{J\in\B:\,J\cap L\neq\varnothing}\beta_J$ as in Proposition~\ref{prop:GammaPareto_Minimum}.
\end{proposition}

Comparing Propositions~\ref{prop:GammaPareto_Minimum} and~\ref{prop:GammaPareto_singular}, we find $X_i\mid\Theta_k^L\overset{d}{=}T^{L}$ for all $i\in L$. This is no coincidence: the proof in Appendix~\ref{app:proofs_sec2} shows that the minimum $T^{L}$ is independent of $\Theta_k^L$ and coincides with every $X_i$, $i\in L$, on $\Theta_k^L$.

Proposition~\ref{prop:GammaPareto_singular} identifies singular mass on the within-class diagonals
\[
\Theta:=\bigcup_{\substack{k\in M\\|N_k|\ge2}}\ \bigcup_{\substack{i,j\in N_k\\i\neq j}}\left\{\xx\in[0,\infty)^n:x_i=x_j\right\}.
\]
To ensure that no singular mass arises elsewhere, we impose a non-degeneracy condition on the joint distribution of the denominators of classes affected by exponential risk factors in $\B_{+}$. Let $M_{\B_{+}}=\bigcup_{J\in\B_{+}}M_J$. We say that the factor systems are \emph{non-degenerate} if
\begin{equation}
\label{eq:ND}
\text{the vectors }\left(\mathds{1}_{\{I\in\AA_k\}}\right)_{I\in\A}\in\R^{|\A|},\ k\in M_{\B_{+}},\text{ are linearly independent},
\end{equation}
or, equivalently, if $(W^{(k)})_{k\in M_{\B_{+}}}$ admits a joint density on $(0,\infty)^{|M_{\B_{+}}|}$. Condition \eqref{eq:ND} holds trivially if $\B_{+}=\varnothing$ or $|M_{\B_{+}}|\le2$, because two distinct non-zero $0$--$1$ vectors are linearly independent. It also holds if every class affected by an exponential risk factor in $\B_+$ possesses a gamma risk factor affecting precisely that class, i.e., $N_k\in\A$ for all $k\in M_{\B_{+}}$. In particular, it is satisfied in the bivariate case, by \eqref{eq:6.2.8arnold} and \eqref{eq:6.2.3arnold}, and by the motivating example \eqref{eq:motiv_model}.

\begin{proposition}
\label{prop:GammaPareto_ACvsSingular}
Let $\PP_{\XX}=\PP_{\ac}+\PP_{\co}$ be the Lebesgue decomposition of the law of $\XX$ into its absolutely continuous and singular parts. For $k\in M$, define $\BB_{k}:=\bigcup_{i\in N_k}\B_i=\{J\in\B:J\cap N_k\neq\varnothing\}$, the set of labels of exponential risk factors affecting the class $N_k$. We say that $N_k$ admits a \emph{separating order} if there is an enumeration $i_1,\dots,i_{|N_k|}$ of $N_k$ such that, writing $U_t=\{i_t,\dots,i_{|N_k|}\}$, for every $t=1,\dots,|N_k|-1$ some $J\in\BB_{k}$ satisfies $J\cap U_t=\{i_t\}$. We call any such label $J$ a \emph{witness} for the $t$-th step of the separating order. If \eqref{eq:ND} holds, then $\PP_{\co}=\PP_{\XX}(\,\cdot\cap\Theta)$ and $\PP_{\ac}=\PP_{\XX}(\,\cdot\setminus\Theta)$. Moreover:
\begin{enumerate}
\item[(i)] \textbf{(Singular part exists)} $\PP_{\co}\neq0$ $\iff$ $\exists\,k\in M,\ i\neq j\in N_k$ with $\B_i\cap\B_j\neq\varnothing$.
\item[(ii)] \textbf{(Purely absolutely continuous)} $\PP_{\co}=0$ $\iff$ for every $k\in M$ with $|N_k|\ge2$ and every $i\neq j\in N_k$, $\B_i\cap\B_j=\varnothing$.
\item[(iii)] \textbf{(Absolutely continuous part exists)} $\PP_{\ac}\neq0$ $\iff$ every $N_k$, $k\in M$, with $|N_k|\ge2$ admits a separating order.
\item[(iv)] \textbf{(Purely singular)} $\PP_{\ac}=0$ $\iff$ some $N_k$, $k\in M$, with $|N_k|\ge2$ admits no separating order.
\end{enumerate}
The implication ``$\Leftarrow$'' in (i) and the implication ``$\Rightarrow$'' in (iii) hold without \eqref{eq:ND}.
\end{proposition}

Since any $J\in\B_{+}$ intersecting $N_k$ contains the entire class, the witnesses $J$ in a separating order necessarily belong to $\B_{-}$. Absolutely continuous behaviour within a class is therefore generated by exponential risk factors confined to that class, whereas factors in $\B_+$ can only create ties within it. Under \eqref{eq:ND}, the singular part is carried entirely by the within-class diagonals and is thus of comonotonic type, in line with Proposition~\ref{prop:GammaPareto_singular}. The next remark shows that \eqref{eq:ND} cannot be dispensed with.

\begin{remark}
\label{rm:GammaPareto_ND}
Let $n=3$, $\A=\{\{1,3\},\{2,3\}\}$ and $\B=\{N\}$. Then $\A_1$, $\A_2$ and $\A_3$ are pairwise distinct, so that every class is a singleton and $\Theta=\varnothing$. However, $W_3=W_1+W_2$ and all components share the systemic exponential risk factor $V=E_N$, whence
\[
\frac{1}{X_3}=\frac{W_1+W_2}{V}=\frac{1}{X_1}+\frac{1}{X_2}\qquad\text{almost surely}.
\]
The law of $\XX$ is therefore concentrated on a two-dimensional surface and is purely singular, although no two components tie. Here $M_{\B_{+}}=\{1,2,3\}$, and \eqref{eq:ND} fails because $(1,1)=(1,0)+(0,1)$. Systemic exponential background risk factors combined with linearly dependent gamma denominators thus generate singular mass.
\end{remark}

\subsection{Special cases and illustrations}
\label{subsec:examples}

The preceding results show how the factor systems $\A$ and $\B$ govern the dependence structure of the Gamma--Pareto sub-mixture. We now collect several notable special cases obtained through specific configurations of these systems.

\begin{example}
\label{ex:GammaPareto_Examples}
Fix any label set $N$ and write $\alpha_i=\alpha_{\{i\}}$ and $\beta_i=\beta_{\{i\}}$.
\begin{itemize}
\item[(1)] By Proposition~\ref{prop:GammaPareto_Independence}, if $\A=\B=\{\{1\},\dots,\{n\}\}$, the components of $\XX$ are mutually independent and
\[
\SF_{\XX}(\xx)=\prod_{i\in N}\left(1+\beta_i x_i\right)^{-\alpha_i}.
\]
\item[(2)] If $N\in\A$ or $N\in\B$, then, by Proposition~\ref{prop:GammaPareto_Independence}, no two distinct components of $\XX$ are independent.
\item[(3)] If $\A=\B=\{N\}$, then $X_1=\cdots=X_n=E_N/G_N$ almost surely, so that $\XX$ is fully comonotonic with
\[
\SF_{\XX}(\xx)=\left(1+\beta_N\max_{i\in N}(x_i)\right)^{-\alpha_N}.
\]
\item[(4)] If $\B_{+}=\varnothing$, the model reduces, by Remark~\ref{rm:GammaPareto_BplusEmpty}, to the hybrid distribution \eqref{eq:BplusEmpty_GammaParetoHybrid}. The distribution \eqref{eq:6.2.8arnold} is recovered by further setting $\B=\{\{1\},\dots,\{n\}\}$, and \eqref{eq:6.2.3arnold} by setting $\A=\{N\}$.
\item[(5)] If $\B_{-}\subseteq\{N_1,\dots,N_m\}$, in particular if $\B_{-}=\varnothing$, and some class satisfies $|N_k|\ge2$, then $N_k$ admits no separating order, because no exponential risk factors affecting $N_k$ can meet $N_k$ in a single point. By Proposition~\ref{prop:GammaPareto_ACvsSingular}(iv), which does not require \eqref{eq:ND} in this direction, $\PP_{\ac}=0$ and the law of $\XX$ is purely singular. Its survival function remains of the form \eqref{eq:GammaPareto}; for $\B_{-}=\varnothing$,
\[
\Lambda_I^{\pi}(\xx)=\sum_{\substack{J\in\B_{+}(\xx)\\k_J^{\pi}\in M_I}}\beta_J\,x^{*}_{k_J^{\pi}}.
\]
\item[(6)] If $\{\{1\},\dots,\{n\}\}\subseteq\A$, every class is a singleton with an idiosyncratic gamma risk factor. Then \eqref{eq:ND} holds, $\Theta=\varnothing$, and Proposition~\ref{prop:GammaPareto_ACvsSingular}(ii) yields $\PP_{\co}=0$: the law of $\XX$ is purely absolutely continuous. Identifying each class with its unique element, \eqref{eq:Lambda_MP} simplifies to
\[
\Lambda_I^{\pi}(\xx)=\sum_{\substack{J\in\B_{-}\\k_J\in M_I}}\beta_J\,x_{k_J}+\sum_{\substack{J\in\B_{+}(\xx)\\k_J^{\pi}\in M_I}}\beta_J\,x_{k_J^{\pi}}.
\]
\item[(7)] The motivating example corresponds to $\A=\{\{1\},\ldots,\{n\}\}$ and $\B=\{N\}$, a special case of (6). Here $\Lambda_{\{i\}}^{\pi}(\xx)=\beta_Nx_i\mathds{1}_{\{\pi(1)=i\}}$ for $\xx\in(0,\infty)^n$, so that the component associated with $\pi$ is $\SF_{X_{\pi(1)}}(x_{\pi(1)})$. Summing the weights of the $(n-1)!$ orderings with $\pi(1)=i$ gives $p_i(\xx)$, and Theorem~\ref{thm:mainPD} reduces to \eqref{eq:sub-m}.
\end{itemize}
\end{example}

As illustrated by Example~\ref{ex:GammaPareto_Examples}, the Gamma--Pareto sub-mixture family encompasses a broad spectrum of heavy-tailed dependence structures, ranging from complete independence to full comonotonicity, together with numerous intermediate configurations. To further elucidate these features, we conclude this section with a detailed study of the bivariate case, which provides a particularly transparent setting in which the principal mechanisms of the general construction can be examined explicitly.

\begin{example}
\label{ex:GammaPareto_Bivariate}
Let $N=\{1,2\}$ and $\XX=(X_1,X_2)=(V_1/W_1,V_2/W_2)$. We build five cases from the same library of ingredients, switching individual factors on and off:
\begin{itemize}
\item idiosyncratic gamma risk factors indexed by $I_1=\{1\}$, $I_2=\{2\}$, with shapes $\alpha_1,\alpha_2$;
\item idiosyncratic exponential risk factors indexed by  $J_1=\{1\}$, $J_2=\{2\}$, with rates $\beta_1,\beta_2$;
\item an optional systemic gamma risk factor indexed by $I_{12}=N$ with shape $\alpha_{12}$;
\item an optional systemic exponential risk factor indexed by $J_{12}=N$ with rate $\beta_{12}$.
\end{itemize}
We write $G_1,G_2,G_{12}$ for the independent unit-rate gamma variables associated with $I_1,I_2,I_{12}$, and $E_1,E_2,E_{12}$ for the independent exponentials associated with $J_1,J_2,J_{12}$. In Cases (1)--(4) the idiosyncratic gamma risk factors are retained, so that $\A_1\neq\A_2$ regardless of whether $I_{12}$ is included; hence $m=2$, $N_1=\{1\}$ and $N_2=\{2\}$. Since \eqref{eq:ND} holds automatically in the bivariate case and $\Theta=\varnothing$, Proposition~\ref{prop:GammaPareto_ACvsSingular}(ii) shows that the laws in Cases (1)--(4) are purely absolutely continuous. Moreover, by Proposition~\ref{prop:GammaPareto_Independence}, independence holds in Case (1) only.

\medskip
\noindent\emph{Case (1): $\A_1\cap\A_2=\varnothing$, $\B_1\cap\B_2=\varnothing$.} Here $\A=\{I_1,I_2\}$, $\B=\{J_1,J_2\}$, $W_i=G_i$ and $V_i=E_i$, $i=1,2$, and $\B_{+}=\varnothing$, which yields the independence survival function
\[
\SF_{\XX}(x_1,x_2)=(1+\beta_1x_1)^{-\alpha_1}(1+\beta_2x_2)^{-\alpha_2}.
\]

\medskip
\noindent\emph{Case (2): $\A_1\cap\A_2=\varnothing$, $\B_1\cap\B_2\neq\varnothing$.} Here $\A=\{I_1,I_2\}$ and $\B=\{J_1,J_2,J_{12}\}$, so that $W_1=G_1$ and $W_2=G_2$ remain independent, whereas $V_1=\min(E_1,E_{12})$ and $V_2=\min(E_2,E_{12})$ form a bivariate Marshall--Olkin pair. Since $M_{J_{12}}=\{1,2\}$, we have $J_{12}\in\B_{+}$, and
\begin{align*}
\SF_{\XX}(x_1,x_2)&=p^{(1,2)}\,\underbrace{(1+(\beta_1+\beta_{12})x_1)^{-\alpha_1}(1+\beta_2x_2)^{-\alpha_2}}_{\pi=(1,2)}\\
&\quad+\,p^{(2,1)}\,\underbrace{(1+\beta_1x_1)^{-\alpha_1}(1+(\beta_2+\beta_{12})x_2)^{-\alpha_2}}_{\pi=(2,1)},
\end{align*}
where $p^{\pi}=\PP(\GG\in\Omega^{\pi}(x_1,x_2))$. As $\Lambda_{I_1}^{(1,2)}=(\beta_1+\beta_{12})x_1$ and $\Lambda_{I_2}^{(1,2)}=\beta_2x_2$, we have $\Omega^{(1,2)}=\{c_1g_1\ge c_2g_2\}$ with $c_1=x_1/(1+(\beta_1+\beta_{12})x_1)$ and $c_2=x_2/(1+\beta_2x_2)$. Using the Dirichlet representation, i.e., $G_1/(G_1+G_2)\sim\mathrm{Beta}(\alpha_1,\alpha_2)$, we obtain
\[
p^{(1,2)}=\PP(c_1G_1\ge c_2G_2)=\bar I_{c_2/(c_1+c_2)}(\alpha_1,\alpha_2),\qquad \bar I_t(a,b):=1-I_t(a,b),
\]
where $I_t(a,b)$ denotes the regularized incomplete beta function. Symmetrically, with $c_1'=x_1/(1+\beta_1x_1)$ and $c_2'=x_2/(1+(\beta_2+\beta_{12})x_2)$, we have $p^{(2,1)}=\bar I_{c_1'/(c_1'+c_2')}(\alpha_2,\alpha_1)$. Unlike in Case (1), $\Lambda^{\pi}$ now depends on $\pi$ through the class to which the contribution of the systemic exponential risk factor $E_{12}$ is assigned. The cones $\Omega^{(1,2)}$ and $\Omega^{(2,1)}$ are therefore rescaled differently and do not cover $[0,\infty)^2$. For instance, for $\alpha_1=\alpha_2=\beta_1=\beta_2=\beta_{12}=x_1=x_2=1$ we have $c_1=1/3$ and $c_2=1/2$, whence $p^{(1,2)}=p^{(2,1)}=2/5$ and a deficit of $1/5$. Removing the idiosyncratic exponential risk factors, i.e., taking $\B=\{J_{12}\}$, recovers the bivariate version of the motivating example.

\medskip
\noindent\emph{Case (3): $\A_1\cap\A_2\neq\varnothing$, $\B_1\cap\B_2=\varnothing$.} Here $\A=\{I_1,I_2,I_{12}\}$ and $\B=\{J_1,J_2\}$: $W_1=G_1+G_{12}$ and $W_2=G_2+G_{12}$ share the factor $G_{12}$, while $V_1=E_1$ and $V_2=E_2$ remain idiosyncratic. Since $\B_{+}=\varnothing$, $\Lambda^{\pi}$ does not depend on $\pi$, the two weights sum to one, and the representation collapses to the bivariate Arnold-type model in \eqref{eq:6.2.8arnold},
\[
\SF_{\XX}(x_1,x_2)=(1+\beta_1x_1)^{-\alpha_1}(1+\beta_2x_2)^{-\alpha_2}(1+\beta_1x_1+\beta_2x_2)^{-\alpha_{12}}.
\]

\medskip
\noindent\emph{Case (4): $\A_1\cap\A_2\neq\varnothing$, $\B_1\cap\B_2\neq\varnothing$.} Here $\A=\{I_1,I_2,I_{12}\}$ and $\B=\{J_1,J_2,J_{12}\}$, and both mechanisms operate simultaneously: $W_1=G_1+G_{12}$ and $W_2=G_2+G_{12}$ share a gamma factor, while $V_1=\min(E_1,E_{12})$ and $V_2=\min(E_2,E_{12})$ share the exponential risk factor indexed by $J_{12}\in\B_{+}$. Writing $\sol{\beta}_1=\beta_1+\beta_{12}$ and $\sol{\beta}_2=\beta_2+\beta_{12}$ for the aggregate rates of Section~\ref{subsec:structural}, we obtain
\begin{align*}
\SF_{\XX}(x_1,x_2)&=\PP\bigl(\GG\in\Omega^{(1,2)}\bigr)\,(1+\sol{\beta}_1x_1)^{-\alpha_1}(1+\beta_2x_2)^{-\alpha_2}\bigl(1+\sol{\beta}_1x_1+\beta_2x_2\bigr)^{-\alpha_{12}}\\
&\quad+\PP\bigl(\GG\in\Omega^{(2,1)}\bigr)\,(1+\beta_1x_1)^{-\alpha_1}(1+\sol{\beta}_2x_2)^{-\alpha_2}\bigl(1+\beta_1x_1+\sol{\beta}_2x_2\bigr)^{-\alpha_{12}},
\end{align*}
where
\[
\Omega^{(1,2)}=\Bigl\{x_1\Bigl(\tfrac{g_1}{1+\sol{\beta}_1x_1}+\tfrac{g_{12}}{1+\sol{\beta}_1x_1+\beta_2x_2}\Bigr)\ge x_2\Bigl(\tfrac{g_2}{1+\beta_2x_2}+\tfrac{g_{12}}{1+\sol{\beta}_1x_1+\beta_2x_2}\Bigr)\Bigr\},
\]
and $\Omega^{(2,1)}$ is obtained by interchanging the roles of the indices $1$ and $2$. In contrast with Case (2), the cones now constrain three independent gamma variables, since $G_{12}$ enters both sides of the defining inequality. Conditioning on $G_{12}$ reduces each weight to a one-dimensional integral of the distribution function of a weighted difference of two independent gamma variables, but no single incomplete beta evaluation is available in general. This illustrates how the simultaneous activation of both dependence channels enriches the structure of the weights.

\medskip
\noindent\emph{Case (5): singular components.} In Cases (1)--(4), $\A_1\neq\A_2$ because the idiosyncratic gamma risk factors are always retained; every class is therefore a singleton, and no singular mass can arise. Singularity requires two indices in the same class, i.e., $\A_1=\A_2$. We thus drop the idiosyncratic gamma risk factors, so that $\A=\{I_{12}\}$, $m=1$ and $N_1=\{1,2\}$. With $\B=\{J_1,J_2,J_{12}\}$ as in Case (4), the exponential risk factor indexed by $J_{12}=N_1$ now belongs to $\B_{-}$, and
\[
\SF_{\XX}(x_1,x_2)=\left(1+\beta_1x_1+\beta_2x_2+\beta_{12}\max(x_1,x_2)\right)^{-\alpha_{12}},
\]
which is the bivariate version of \eqref{eq:6.2.3arnold}. Proposition~\ref{prop:GammaPareto_singular} gives
\[
\PP(X_1=X_2)=\PP\left(\Theta_1^{\{1,2\}}\right)=\frac{\beta_{12}}{\beta_1+\beta_2+\beta_{12}}>0,
\]
and $X_1\mid\Theta_1^{\{1,2\}}$ is Lomax distributed with shape $\alpha_{12}$ and rate $\beta_1+\beta_2+\beta_{12}$. The law has both singular and absolutely continuous parts, because the enumeration $(1,2)$ with witness $J_1$ is separating. Removing both idiosyncratic exponential risk factors, i.e., taking $\B=\{J_{12}\}$, yields the fully comonotonic law of Example~\ref{ex:GammaPareto_Examples}(3).
\end{example}
\section{Dependence Measures}
\label{sec:Dependence_Measures}

Section~\ref{sec:GammaPareto} established the principal distributional properties of the Gamma--Pareto sub-mixture, including independence and the presence of singular and absolutely continuous parts. We now quantify \emph{how strongly} the risk components depend on each other. Positive dependence plays a fundamental role in actuarial science, affecting the frequency of simultaneous large risk realizations, aggregate risk, diversification benefits, capital allocation and solvency assessment. We first establish global positive dependence properties and the covariance structure of the model in Sections~\ref{subsec:association} and~\ref{subsec:covariance}, and then turn to extremal dependence in Section~\ref{subsec:extremal}. A recurring theme emerges: the gamma risk factors determine the presence of asymptotic tail dependence and the decay rates of joint exceedance probabilities, whereas the exponential risk factors affect the associated tail constants.

\subsection{Association and orthant dependence}
\label{subsec:association}

A random vector $\YY=(Y_1,\dots,Y_p)$ is said to be \emph{associated} if
\[
\Cov\!\left(\psi_1(\YY),\psi_2(\YY)\right)\ge0
\]
for every pair of coordinate-wise non-decreasing functions $\psi_1,\psi_2:\R^p\rightarrow\R$ for which the covariance exists. Association was introduced by \citet{Esaryetal1967} and underpins numerous positive dependence concepts used in probability, statistics, reliability and insurance. We shall use the following standard properties \citep{Esaryetal1967}.

\smallskip
(P1)--\emph{Closure under monotone transformations.} If $\YY$ is associated and $\xi_1,\ldots,\xi_k:\R^p\rightarrow\R$ are coordinate-wise non-decreasing, then $\left(\xi_1(\YY),\ldots,\xi_k(\YY)\right)$ is associated.

\smallskip
(P2)--\emph{Independence implies association.} Every random vector with independent components is associated.

\begin{proposition}
\label{prop:GammaPareto_Associated}
The Gamma--Pareto sub-mixture random vector $\XX$ is associated.
\end{proposition}

\begin{proof}
Consider the vector $\ZZ=\bigl((E_J)_{J\in\B},\,(-G_I)_{I\in\A}\bigr)$, whose components are independent and which is therefore associated by (P2). For every $i\in N$, the component
\[
X_i=\frac{\min_{J\in\B_i}E_J}{\sum_{I\in\A_i}G_I}
\]
is non-decreasing in each $E_J$ and non-increasing in each $G_I$, i.e., it is a coordinate-wise non-decreasing function of $\ZZ$. Property (P1) yields the association of $\XX$.
\end{proof}

The same argument shows that $\VV$, being a non-decreasing function of $(E_J)_{J\in\B}$, and $\UU=(1/W_1,\ldots,1/W_n)$, being a non-decreasing function of $(-G_I)_{I\in\A}$, are associated; this fact is used in Proposition~\ref{prop:GammaPareto_covdecomp} below.

\begin{corollary}
\label{cor:GammaPareto_PUODPLODCov}
The Gamma--Pareto sub-mixture random vector $\XX$ satisfies the following classical positive dependence properties \citep{Esaryetal1967}.
\begin{enumerate}
\item[(a)] (\emph{Positive upper orthant dependence, PUOD}) $\displaystyle \SF_{\XX}(x_1,\ldots,x_n)\ge\prod_{i=1}^n\SF_{X_i}(x_i)$ for all $\xx\in[0,\infty)^n$.
\item[(b)] (\emph{Positive lower orthant dependence, PLOD}) $\displaystyle F_{\XX}(x_1,\ldots,x_n)\ge\prod_{i=1}^n F_{X_i}(x_i)$ for all $\xx\in[0,\infty)^n$, where $F$ denotes the cumulative distribution function.
\item[(c)] Taking $\psi_1(\xx)=x_i$ and $\psi_2(\xx)=x_j$ in the definition of association gives $\Cov(X_i,X_j)\ge0$ for all $i,j\in N$ for which the covariance exists.
\end{enumerate}
Here the margins are the Lomax distributions with shapes $\sol{\alpha}_i$ and rates $\sol{\beta}_i$ identified in Section~\ref{subsec:structural}, and all three properties hold without any restriction on the factor systems $\A$ and $\B$.
\end{corollary}

Association is among the strongest and most useful notions of positive dependence in actuarial modelling, as it implies several weaker dependence concepts while remaining analytically tractable. PUOD guarantees that joint exceedances occur at least as frequently as under independence, which is particularly relevant for simultaneous large risk realizations and stress scenarios affecting several risk components. PLOD makes the analogous statement for the joint occurrence of small and moderate risk realizations, which is informative for reserving, credibility and aggregate claims modelling. Non-negative covariances, in turn, imply that dependence cannot be offset through negatively correlated business lines, so that the model provides conservative dependence structures for capital allocation and enterprise risk management. We refer to \citet{Esaryetal1967} for the probabilistic theory of association and to \citet{JoagDev1983} for its negative counterpart, to \citet{Dhaene2002} and \citet{Denuit2005} for its actuarial applications, and to \citet{McNeil2015} for a comprehensive treatment of dependence modelling in quantitative risk management.

\subsection{Covariance structure}
\label{subsec:covariance}

Owing to the explicit stochastic construction of the model, its pairwise covariances admit a transparent decomposition into the contributions of the Marshall--Olkin and additive gamma risk factor structures and their interaction. Throughout this subsection, we use the aggregate shapes $\sol{\alpha}_i$ and $\sol{\alpha}_{ij}$ and the aggregate rates $\sol{\beta}_i$ and $\sol{\beta}_{ij}$ introduced in Section~\ref{subsec:structural}.

\begin{proposition}
\label{prop:GammaPareto_covdecomp}
Let $i\neq j\in N$ and assume that $\sol{\alpha}_i>1$, $\sol{\alpha}_j>1$ and $\sol{\alpha}_i+\sol{\alpha}_j-\sol{\alpha}_{ij}>2$. Define
\[
a_i:=\e\!\left[\frac1{W_i}\right]=\frac1{\sol{\alpha}_i-1},\qquad
b_i:=\e[V_i]=\frac1{\sol{\beta}_i},
\]
and analogously $a_j$ and $b_j$, and let
\[
\C_W:=\Cov\!\left(\frac1{W_i},\frac1{W_j}\right),\qquad
\C_V:=\Cov(V_i,V_j).
\]
Then $\Cov(X_i,X_j)$ is finite and
\begin{equation}
\label{eq:GammaPareto_CovarianceDecomposition}
\Cov(X_i,X_j)=\C_V\C_W+a_ia_j\,\C_V+b_ib_j\,\C_W,
\end{equation}
where $\C_V\ge0$ and $\C_W\ge0$. The moment conditions are also necessary for the finiteness of $\Cov(X_i,X_j)$.
\end{proposition}

\begin{proof}
Since $\VV$ is independent of $\WW$, we have $\e[X_iX_j]=\e[V_iV_j]\,\e[1/(W_iW_j)]=(\C_V+b_ib_j)(\C_W+a_ia_j)$, where $\e[1/(W_iW_j)]$ is finite if and only if the stated conditions hold (see the proof of Corollary~\ref{cor:GammaPareto_Covariances}). Subtracting $\e[X_i]\,\e[X_j]=a_ib_ia_jb_j$ yields \eqref{eq:GammaPareto_CovarianceDecomposition}. Finally, $\VV$ and $\UU=1/\WW$ are associated, as noted after Proposition~\ref{prop:GammaPareto_Associated}, whence $\C_V\ge0$ and $\C_W\ge0$.
\end{proof}

The decomposition \eqref{eq:GammaPareto_CovarianceDecomposition} is particularly informative from a modelling perspective. It separates the dependence induced by the Marshall--Olkin risk factor structure, represented by $\C_V$, from that generated by the additive gamma risk factor structure, represented by $\C_W$. The first term captures dependence arising through the systemic exponential background risk factors encoded by $\B$, whereas the second reflects dependence generated by the shared gamma factors encoded by $\A$. The covariance therefore consists of three distinct contributions: a purely Marshall--Olkin term $a_ia_j\C_V$, a purely additive gamma term $b_ib_j\C_W$, and an interaction term $\C_V\C_W$ accounting for the simultaneous presence of both mechanisms. As a consistency check, for $j=i$ the same computation, with $\C_V=\Var(V_i)=\sol{\beta}_i^{-2}$ and $\C_W=\Var(1/W_i)$, returns the Lomax variance of Section~\ref{subsec:structural}.

\begin{corollary}
\label{cor:GammaPareto_Covariances}
For $i\neq j\in N$, the Marshall--Olkin contribution is
\[
\C_V=\frac{\sol{\beta}_{ij}}{\sol{\beta}_i\,\sol{\beta}_j\,(\sol{\beta}_i+\sol{\beta}_j-\sol{\beta}_{ij})},
\]
so that $\C_V=0$ if and only if $\B_i\cap\B_j=\varnothing$. Under the conditions of Proposition~\ref{prop:GammaPareto_covdecomp}, the additive gamma contribution is
\begin{equation}
\label{eq:GammaPareto_CW}
\C_W=\frac{{}_3F_2\!\left(1,1,\sol{\alpha}_{ij};\sol{\alpha}_i,\sol{\alpha}_j;1\right)-1}{(\sol{\alpha}_i-1)(\sol{\alpha}_j-1)}
=\frac{1}{(\sol{\alpha}_i-1)(\sol{\alpha}_j-1)}\sum_{r=1}^{\infty}\frac{r!\,(\sol{\alpha}_{ij})_r}{(\sol{\alpha}_i)_r(\sol{\alpha}_j)_r},
\end{equation}
where ${}_3F_2$ denotes the generalized hypergeometric function and $(c)_r=c(c+1)\cdots(c+r-1)$ the Pochhammer symbol. In particular, $\C_W=0$ if and only if $\A_i\cap\A_j=\varnothing$. Two important special cases admit elementary closed forms.
\begin{itemize}
\item If $\A_i=\A_j$, i.e., $i$ and $j$ belong to the same class $N_k$, then $W_i=W_j=W^{(k)}$, $\sol{\alpha}_i=\sol{\alpha}_j=\sol{\alpha}_{ij}=\sol{\alpha}^{(k)}$, and
\[
\C_W=\Var\!\left(\frac1{W^{(k)}}\right)=\frac1{(\sol{\alpha}^{(k)}-1)^2(\sol{\alpha}^{(k)}-2)},\qquad\sol{\alpha}^{(k)}>2.
\]
\item If $\A_j\subsetneq\A_i$, so that $\sol{\alpha}_{ij}=\sol{\alpha}_j$, then
\[
\C_W=\frac1{(\sol{\alpha}_{j}-1)(\sol{\alpha}_i-1)(\sol{\alpha}_i-2)},\qquad\sol{\alpha}_{j}>1,\ \sol{\alpha}_i>2,
\]
and symmetrically if $\A_i\subsetneq\A_j$.
\end{itemize}
\end{corollary}

\begin{proof}
Write $\nu_i=\sol{\beta}_i-\sol{\beta}_{ij}$, $\nu_j=\sol{\beta}_j-\sol{\beta}_{ij}$ and $\nu=\sol{\beta}_i+\sol{\beta}_j-\sol{\beta}_{ij}$. By Proposition~\ref{prop:GammaPareto_marginalization}, $\PP(V_i>s_1,V_j>s_2)=\exp\{-\nu_is_1-\nu_js_2-\sol{\beta}_{ij}\max(s_1,s_2)\}$, and splitting the integral at $s_1=s_2$ gives
\[
\e[V_iV_j]=\int_0^\infty\!\!\int_0^\infty\PP(V_i>s_1,V_j>s_2)\,\d s_1\,\d s_2=\frac{1}{\sol{\beta}_j\nu}+\frac{1}{\sol{\beta}_i\nu}=\frac{\sol{\beta}_i+\sol{\beta}_j}{\sol{\beta}_i\sol{\beta}_j\nu}.
\]
Subtracting $b_ib_j=(\sol{\beta}_i\sol{\beta}_j)^{-1}$ yields $\C_V$. For $\C_W$, write $W_i=\zeta_i+\zeta_0$ and $W_j=\zeta_j+\zeta_0$, where $\zeta_0=\sum_{I\in\A_i\cap\A_j}G_I$, $\zeta_i=\sum_{I\in\A_i\setminus\A_j}G_I$ and $\zeta_j=\sum_{I\in\A_j\setminus\A_i}G_I$ are independent gamma variables with shapes $\kappa_0=\sol{\alpha}_{ij}$, $\kappa_i=\sol{\alpha}_i-\sol{\alpha}_{ij}$ and $\kappa_j=\sol{\alpha}_j-\sol{\alpha}_{ij}$, a gamma variable with zero shape being identically zero. Using $w^{-1}=\int_0^\infty e^{-sw}\d s$, Tonelli's theorem and the substitutions $y=(1+s_1)^{-1}$, $z=(1+s_2)^{-1}$, we obtain
\[
\e\!\left[\frac{1}{W_iW_j}\right]=\int_0^\infty\!\!\int_0^\infty\frac{\d s_1\,\d s_2}{(1+s_1)^{\kappa_i}(1+s_2)^{\kappa_j}(1+s_1+s_2)^{\kappa_0}}
=\int_0^1\!\!\int_0^1\frac{y^{\sol{\alpha}_i-2}\,z^{\sol{\alpha}_j-2}}{\bigl(1-(1-y)(1-z)\bigr)^{\kappa_0}}\,\d y\,\d z.
\]
Expanding $(1-(1-y)(1-z))^{-\kappa_0}=\sum_{r\ge0}(\kappa_0)_r(1-y)^r(1-z)^r/r!$ and integrating termwise with Beta integrals, which converge since $\sol{\alpha}_i>1$ and $\sol{\alpha}_j>1$, gives
\[
\e\!\left[\frac{1}{W_iW_j}\right]=\frac{1}{(\sol{\alpha}_i-1)(\sol{\alpha}_j-1)}\sum_{r=0}^{\infty}\frac{r!\,(\sol{\alpha}_{ij})_r}{(\sol{\alpha}_i)_r(\sol{\alpha}_j)_r}.
\]
The series has positive terms of order $r^{-(\sol{\alpha}_i+\sol{\alpha}_j-\sol{\alpha}_{ij}-1)}$ and therefore converges if and only if $\sol{\alpha}_i+\sol{\alpha}_j-\sol{\alpha}_{ij}>2$. Subtracting $a_ia_j$, i.e., the term $r=0$, yields \eqref{eq:GammaPareto_CW}, and all terms with $r\ge1$ vanish if and only if $\sol{\alpha}_{ij}=0$. The two special cases follow from Gauss's summation formula ${}_2F_1(1,1;d;1)=(d-1)/(d-2)$, $d>2$, since the ${}_3F_2$ reduces to ${}_2F_1(1,1;\sol{\alpha}^{(k)};1)$ and ${}_2F_1(1,1;\sol{\alpha}_i;1)$, respectively.
\end{proof}

\begin{example}
\label{ex:GammaPareto_BivariateCovariance}
Consider the bivariate models of Example~\ref{ex:GammaPareto_Bivariate}. Applying \eqref{eq:GammaPareto_CovarianceDecomposition} together with Corollary~\ref{cor:GammaPareto_Covariances} yields explicit covariances for each dependence configuration.

\medskip
\noindent\emph{Case (1).} Here $\C_W=\C_V=0$, and $\Cov(X_1,X_2)=0$, in agreement with the independence of $X_1$ and $X_2$.

\medskip
\noindent\emph{Case (2).} Here $\A_1\cap\A_2=\varnothing$, so that $\C_W=0$, while $\sol{\alpha}_1=\alpha_1$, $\sol{\alpha}_2=\alpha_2$, $\sol{\beta}_1=\beta_1+\beta_{12}$, $\sol{\beta}_2=\beta_2+\beta_{12}$ and $\sol{\beta}_{12}=\beta_{12}$. Only the Marshall--Olkin risk factor structure contributes, and for $\alpha_1,\alpha_2>1$,
\[
\Cov(X_1,X_2)=a_1a_2\,\C_V=\frac1{(\alpha_1-1)(\alpha_2-1)}\cdot\frac{\beta_{12}}{(\beta_1+\beta_{12})(\beta_2+\beta_{12})(\beta_1+\beta_2+\beta_{12})}.
\]

\medskip
\noindent\emph{Case (3).} Here $\B_1\cap\B_2=\varnothing$, so that $\C_V=0$, while $\sol{\alpha}_1=\alpha_1+\alpha_{12}$, $\sol{\alpha}_2=\alpha_2+\alpha_{12}$, $\sol{\alpha}_{12}=\alpha_{12}$, $b_1=1/\beta_1$ and $b_2=1/\beta_2$. Only the additive gamma contribution remains:
\[
\Cov(X_1,X_2)=b_1b_2\,\C_W=\frac1{\beta_1\beta_2}\cdot\frac{{}_3F_2(1,1,\alpha_{12};\alpha_1+\alpha_{12},\alpha_2+\alpha_{12};1)-1}{(\alpha_1+\alpha_{12}-1)(\alpha_2+\alpha_{12}-1)},
\]
provided that $\alpha_1+\alpha_{12}>1$, $\alpha_2+\alpha_{12}>1$ and $\alpha_1+\alpha_2+\alpha_{12}>2$.

\medskip
\noindent\emph{Case (4).} Both mechanisms are active: $\C_V$ coincides with that of Case (2), and $\C_W$ with that of Case (3). With $a_i=(\alpha_i+\alpha_{12}-1)^{-1}$ and $b_i=(\beta_i+\beta_{12})^{-1}$, $i=1,2$,
\[
\Cov(X_1,X_2)=\C_V\C_W+\frac{\C_V}{(\alpha_1+\alpha_{12}-1)(\alpha_2+\alpha_{12}-1)}+\frac{\C_W}{(\beta_1+\beta_{12})(\beta_2+\beta_{12})},
\]
and, unlike in the previous cases, all three terms of \eqref{eq:GammaPareto_CovarianceDecomposition} contribute.

\medskip
\noindent\emph{Case (5) and nested variants.} If $\A=\{I_{12}\}$, as in Case (5), then $\sol{\alpha}_1=\sol{\alpha}_2=\sol{\alpha}_{12}=\alpha_{12}$ and $\C_W=(\alpha_{12}-1)^{-2}(\alpha_{12}-2)^{-1}$ for $\alpha_{12}>2$. If only one idiosyncratic gamma risk factor is retained, say $\A=\{I_1,I_{12}\}$, then $\A_2\subsetneq\A_1$, $\sol{\alpha}_2=\sol{\alpha}_{12}=\alpha_{12}$ and $\sol{\alpha}_1=\alpha_1+\alpha_{12}$, whence
\[
\C_W=\frac1{(\alpha_{12}-1)(\alpha_1+\alpha_{12}-1)(\alpha_1+\alpha_{12}-2)},\qquad\alpha_{12}>1,\ \alpha_1+\alpha_{12}>2,
\]
and symmetrically for $\A=\{I_2,I_{12}\}$.

\medskip
This example illustrates how the covariance decomposition cleanly separates the dependence arising from the additive gamma and Marshall--Olkin risk factor structures. Depending on the structure of $\A$ and $\B$, either source of dependence may act alone, both may contribute simultaneously, or the components may be independent.
\end{example}

The corresponding Pearson correlation coefficients follow by normalizing the covariances with the Lomax standard deviations. Their existence requires $\sol{\alpha}_i,\sol{\alpha}_j>2$, which automatically implies the conditions of Proposition~\ref{prop:GammaPareto_covdecomp} since $\sol{\alpha}_i+\sol{\alpha}_j-\sol{\alpha}_{ij}\ge\max(\sol{\alpha}_i,\sol{\alpha}_j)$. For heavy-tailed risks, however, correlation is well known to be an unreliable summary of dependence: it may fail to exist, it depends on the margins, and it says little about the joint behaviour of extreme risk realizations \citep{McNeil2015}. This motivates the extremal dependence measures studied next.

\subsection{Extremal dependence}
\label{subsec:extremal}

Association, covariances and correlations are global measures of dependence, capturing the interactions of the risk components over the whole domain. In risk management, however, the primary concern is often the model's ability to generate simultaneous extreme risks \citep{Embrechts1997,Resnick2007}. We show below that the joint survival function of $\XX$ is regularly varying along every ray, with an index determined entirely by the additive gamma system $\A$ and a limit function to which the Marshall--Olkin system $\B$ contributes only through its rates. This separation yields sharp, easily checked criteria for tail dependence and asymptotic independence between any pair of components.

For $\xx\in[0,\infty)^n\setminus\{\mathbf{0}\}$, let $\A_+(\xx)=\{I\in\A:M_I\cap M_+(\xx)\neq\varnothing\}$ be the set of labels of \emph{active gamma risk factors}, namely
those affecting at least one risk component whose threshold $x_j$
is positive, and let
\[
\sol{\alpha}_{M_+}:=\sum_{I\in\A_+(\xx)}\alpha_I.
\]
For $\uu=(u_I,I\in\A_+(\xx))\in(0,\infty)^{|\A_+(\xx)|}$, write $w_j(\uu)=\sum_{I\in\A_j}u_I$ for every $j$ with $x_j>0$, which is well defined because $\A_j\subseteq\A_+(\xx)$ for such $j$ by \eqref{eq:incidence}, and define
\[
\Phi_{\xx}(\uu)=\sum_{J\in\B}\beta_J\max_{j\in J:\,x_j>0}\bigl(x_j\,w_j(\uu)\bigr),
\]
with the convention that a maximum over the empty set is zero.

\begin{proposition}
\label{prop:GammaPareto_RV}
For every $\xx\in[0,\infty)^n\setminus\{\mathbf{0}\}$,
\begin{equation}
\label{eq:H_integral}
\lim_{t\to\infty}t^{\sol{\alpha}_{M_+}}\,\SF_{\XX}(t\xx)=H(\xx):=\int_{(0,\infty)^{|\A_+(\xx)|}}e^{-\Phi_{\xx}(\uu)}\prod_{I\in\A_+(\xx)}\frac{u_I^{\alpha_I-1}}{\Gamma(\alpha_I)}\,\d\uu\in(0,\infty),
\end{equation}
and $H(s\xx)=s^{-\sol{\alpha}_{M_+}}H(\xx)$ for all $s>0$. If, in addition, $\B_i\cap\B_{-}\neq\varnothing$ for every $i\in N$, then $H$ admits the sub-mixture representation
\[
H(\xx)=\sum_{\pi\in\SS_{M_+}}\PP\bigl(\GG\in\Omega_\infty^\pi(\xx)\bigr)\prod_{I\in\A_+(\xx)}\Lambda_I^\pi(\xx)^{-\alpha_I},
\qquad
\sum_{\pi\in\SS_{M_+}}\PP\bigl(\GG\in\Omega_\infty^\pi(\xx)\bigr)\le1,
\]
where $\Lambda_I^{\pi}$ is given by \eqref{eq:Lambda_MP} and $\Omega_\infty^\pi(\xx)$ is the cone obtained from \eqref{eq:Omega} by replacing each $1+\Lambda_I^{\pi}(\xx)$ with $\Lambda_I^{\pi}(\xx)$.
\end{proposition}

\begin{proof}
By \eqref{eq:app_laplace}, $\SF_{\XX}(t\xx)=\e[\exp\{-t\,\Phi_{\xx}(\GG_+)\}]$, where $\GG_+=(G_I,I\in\A_+(\xx))$: terms with $x_j=0$ vanish, and $W_j=w_j(\GG_+)$ whenever $x_j>0$. Since $\Phi_{\xx}$ is positively homogeneous of degree one, the substitution $\gg=\uu/t$ gives
\[
t^{\sol{\alpha}_{M_+}}\,\SF_{\XX}(t\xx)=\int e^{-\Phi_{\xx}(\uu)}\prod_{I\in\A_+(\xx)}\frac{u_I^{\alpha_I-1}e^{-u_I/t}}{\Gamma(\alpha_I)}\,\d\uu,
\]
and the monotone convergence theorem yields \eqref{eq:H_integral} as $t\to\infty$. The limit is positive because the integrand is positive. To see that it is finite, let $\beta_*=\min_{J\in\B}\beta_J$ and $x_*=\min_{k\in M_+}x_k^{*}>0$. For every active class $k$, choose $j_k\in N_k$ with $x_{j_k}=x_k^{*}$ and an exponential risk factor indexed by $J\in\B_{j_k}$; this gives $\Phi_{\xx}(\uu)\ge\beta_*x_k^{*}\sum_{I\in\AA_k}u_I$. Since every active factor belongs to $\AA_k$ for some $k\in M_+$, averaging over $k\in M_+$ yields $\Phi_{\xx}(\uu)\ge(\beta_*x_*/m_+)\sum_{I\in\A_+(\xx)}u_I$, and the integral is bounded by $\prod_{I\in\A_+(\xx)}(\beta_*x_*/m_+)^{-\alpha_I}<\infty$. Homogeneity of $H$ follows from that of $\Phi_{\xx}$ via the substitution $\uu\mapsto\uu/s$.

Under the additional assumption, fix $I\in\A_+(\xx)$ and $k_0\in M_I\cap M_+$, and let $j^{*}\in N_{k_0}$ satisfy $x_{j^{*}}=x^{*}_{k_0}>0$. For an exponential risk factor indexed by $J\in\B_{j^{*}}\cap\B_{-}$, we have $J\subseteq N_{k_0}$ and $k_J=k_0\in M_I$, so that the factor contributes $\beta_Jx^{*}_{k_0}>0$ to $\Lambda_I^{\pi}(\xx)$ for every $\pi$. Partitioning $(0,\infty)^{|\A_+(\xx)|}$ into the cones \eqref{eq:Omega_tilde} and repeating Step~2 of the proof of Theorem~\ref{thm:mainPD}, we find $\Phi_{\xx}(\uu)=\sum_{I\in\A_+(\xx)}\Lambda_I^{\pi}(\xx)u_I$ on the $\pi$-th cone. The identity $\int_{C}e^{-\sum_I\Lambda_Iu_I}\prod_I u_I^{\alpha_I-1}/\Gamma(\alpha_I)\,\d\uu=\prod_I\Lambda_I^{-\alpha_I}\,\PP\bigl((G_I/\Lambda_I)_I\in C\bigr)$, valid for $\Lambda_I>0$, yields the sub-mixture representation. The inequality follows from Step~4 of that proof, which uses only the positivity of the denominators $1+\Lambda_I^{\pi}$ and therefore applies verbatim to $\Lambda_I^{\pi}$.
\end{proof}

Proposition~\ref{prop:GammaPareto_RV} states that $\SF_{\XX}$ is regularly varying along every ray $\{t\xx:t>0\}$, with a limit function $H$ that is positively homogeneous. The index $\sol{\alpha}_{M_+}$ depends on $\xx$ only through the set of active classes, i.e., through the face of the orthant containing $\xx$, and it is determined by $\A$ alone; $\B$ enters only through $H$. On the axes, $\xx=\ee_i$, the $i$-th standard basis vector of $\R^n$, gives $\sol{\alpha}_{M_+}=\sol{\alpha}_i$ and $H(\ee_i)=\sol{\beta}_i^{-\sol{\alpha}_i}$, recovering the Lomax tail of $X_i$. As more classes become active, the index increases, and $\sol{\alpha}_{M_+}\ge\max_{k\in M_+}\sol{\alpha}^{(k)}$: joint extremes of several classes are rarer than extremes of any single one of them. A classical multivariate regular variation analysis is governed by the heaviest marginal index $\min_{i\in N}\sol{\alpha}_i$ and therefore captures only part of this structure; the successively faster decay on higher-dimensional faces is precisely the phenomenon described by hidden regular variation \citep{Resnick2002,Resnick2007}. Finally, the side condition $\B_i\cap\B_{-}\neq\varnothing$ is needed only for the sub-mixture form of $H$, not for regular variation itself. In the motivating example with $n=2$ and $\alpha_1=\alpha_2=1$, where $\B_{-}=\varnothing$, the survival function computed in Section~\ref{subsec:motivating} gives $t^{2}\SF_{\XX}(t\xx)\to2/(\beta_N^2x_1x_2)$, which is precisely the value of \eqref{eq:H_integral}. More generally, Example~\ref{ex:B_deficit} in Appendix~\ref{app:examples} shows that, in the motivating example, the sub-mixture deficit converges to one at the rate dictated by Proposition~\ref{prop:GammaPareto_RV}, whereas, under the additional assumption of the proposition, the weights converge to the limits $\PP(\GG\in\Omega_\infty^{\pi}(\xx))$, which cannot all vanish since $H(\xx)>0$, so that the deficit stays bounded away from one.

\begin{corollary}
\label{cor:GammaPareto_two-class-RV}
Let $i\neq j\in N$ belong to different classes, $k(i)\neq k(j)$, and let $\xx$ satisfy $x_i,x_j>0$ and $x_l=0$ for $l\notin\{i,j\}$, so that $M_+=\{k(i),k(j)\}$. Then
\[
\sol{\alpha}_{M_+}=\sol{\alpha}_{i}+\sol{\alpha}_{j}-\sol{\alpha}_{ij}\in\bigl[\max(\sol{\alpha}_i,\sol{\alpha}_j),\,\sol{\alpha}_i+\sol{\alpha}_j\bigr].
\]
The upper bound is attained if and only if $\A_i\cap\A_j=\varnothing$, and the lower bound if and only if $\A_i$ and $\A_j$ are nested, i.e., $\A_i\subseteq\A_j$ or $\A_j\subseteq\A_i$.
\end{corollary}

The upper bound is a rate-level counterpart of the PUOD property of Corollary~\ref{cor:GammaPareto_PUODPLODCov}: the joint survival probability decays no faster than the independence rate $t^{-\sol{\alpha}_i-\sol{\alpha}_j}$, and the improvement $\sol{\alpha}_{ij}$ is driven exclusively by the shared gamma factors. Systemic exponential background risk factors do not alter this decay rate.

\begin{corollary}
\label{cor:GammaPareto_lambda}
For any $i\neq j\in N$, the common-threshold tail dependence coefficient
\[
\lambda_{ij}:=\lim_{t\to\infty}\PP(X_j>t\mid X_i>t)
\]
exists and satisfies the following.
\begin{enumerate}
\item[(i)] If $\A_j\not\subseteq\A_i$, then $\lambda_{ij}=0$.
\item[(ii)] If $\A_j\subseteq\A_i$, then, for every $t>0$,
\[
\PP(X_j>t\mid X_i>t)=\left(\frac{1+\sol{\beta}_it}{1+(\sol{\beta}_i+\sol{\beta}_j-\sol{\beta}_{ij})t}\right)^{\sol{\alpha}_j},
\]
and consequently
\begin{equation}
\label{eq:GammaPareto_Lambdaij}
\lambda_{ij}=\left(\frac{\sol{\beta}_i}{\sol{\beta}_i+\sol{\beta}_j-\sol{\beta}_{ij}}\right)^{\sol{\alpha}_j}\in(0,1],
\end{equation}
with $\lambda_{ij}=1$ if and only if $\B_j\subseteq\B_i$.
\end{enumerate}
\end{corollary}

\begin{proof}
(i) If $\A_j\not\subseteq\A_i$, then $i$ and $j$ belong to different classes and $\sol{\alpha}_{ij}<\sol{\alpha}_j$. Proposition~\ref{prop:GammaPareto_RV} and Corollary~\ref{cor:GammaPareto_two-class-RV} give $\SF_{\XX}(t\ee_i+t\ee_j)\sim H(\ee_i+\ee_j)\,t^{-(\sol{\alpha}_i+\sol{\alpha}_j-\sol{\alpha}_{ij})}$ and $\SF_{X_i}(t)\sim\sol{\beta}_i^{-\sol{\alpha}_i}t^{-\sol{\alpha}_i}$, so that their ratio is of order $t^{\sol{\alpha}_{ij}-\sol{\alpha}_j}\to0$. (ii) If $\A_j\subseteq\A_i$, then $W_i=W_j+D$ with $D=\sum_{I\in\A_i\setminus\A_j}G_I\sim\mathrm{Gamma}(\sol{\alpha}_i-\sol{\alpha}_j,1)$ independent of $W_j$, so that $\max(tW_i,tW_j)=tW_i$. By \eqref{eq:app_laplace} applied to the pair $(i,j)$,
\begin{align*}
\SF_{\XX}(t\ee_i+t\ee_j)&=\e\Bigl[e^{-t\left(\sol{\beta}_iW_i+(\sol{\beta}_j-\sol{\beta}_{ij})W_j\right)}\Bigr]
=\e\Bigl[e^{-t\left(\sol{\beta}_iD+(\sol{\beta}_i+\sol{\beta}_j-\sol{\beta}_{ij})W_j\right)}\Bigr]\\
&=\bigl(1+\sol{\beta}_it\bigr)^{-(\sol{\alpha}_i-\sol{\alpha}_j)}\bigl(1+(\sol{\beta}_i+\sol{\beta}_j-\sol{\beta}_{ij})t\bigr)^{-\sol{\alpha}_j}.
\end{align*}
Dividing by $\SF_{X_i}(t)=(1+\sol{\beta}_it)^{-\sol{\alpha}_i}$ yields the exact expression, and letting $t\to\infty$ yields \eqref{eq:GammaPareto_Lambdaij}. Finally, $\lambda_{ij}=1$ if and only if $\sol{\beta}_j=\sol{\beta}_{ij}$, i.e., $\B_j\subseteq\B_i$.
\end{proof}

The coefficient $\lambda_{ij}$ is asymmetric: if $\A_j\subsetneq\A_i$, then $\lambda_{ij}>0$ whereas $\lambda_{ji}=0$. This reflects the fact that $X_j$, which is driven by fewer gamma factors, has the heavier tail, $\sol{\alpha}_j<\sol{\alpha}_i$. An exceedance of $X_i$ requires $W_i$, and hence all gamma factors of $X_j$, to be small, whereas an exceedance of $X_j$ carries no information about the remaining factors $\A_i\setminus\A_j$ of $X_i$. When, in addition, $\B_j\subseteq\B_i$, then $V_j\ge V_i$ and $W_j\le W_i$, so that $X_j\ge X_i$ almost surely and $\lambda_{ij}=1$. Corollary~\ref{cor:GammaPareto_lambda} thus exhibits the separation of roles once more: whether common-threshold tail dependence is present is decided by $\A$, while its magnitude \eqref{eq:GammaPareto_Lambdaij} is governed by the rates of $\B$.

Because $\lambda_{ij}$ compares the components at a common threshold, it mixes dependence with differences in the marginal tails. We therefore also consider the copula-based upper tail dependence coefficient $\chi_{ij}$ \citep{Sibuya1959,Joe1997,Coles1999} and, for asymptotically independent pairs, the coefficient of tail dependence $\eta_{ij}$ of \citet{Ledford1997}, which measures the rate at which joint extremes vanish; see also \citet{Heffernan2000} and the related notion of tail order in \citet{Hua2011a}.

\begin{proposition}
\label{prop:GammaPareto_chi}
Let $i\neq j\in N$, denote by $q_i$ and $q_j$ the quantile functions of $X_i$ and $X_j$, and define
\[
\chi_{ij}:=\lim_{u\to1^{-}}\PP\left(X_j>q_j(u)\mid X_i>q_i(u)\right),
\qquad
\eta_{ij}:=\lim_{u\to1^{-}}\frac{\log(1-u)}{\log\PP\left(X_i>q_i(u),\,X_j>q_j(u)\right)}.
\]
Both limits exist, and:
\begin{enumerate}
\item[(i)] if $i$ and $j$ belong to different classes, then $\chi_{ij}=0$;
\item[(ii)] if $i$ and $j$ belong to the same class, then
\begin{equation}
\label{eq:GammaPareto_Chiij}
\chi_{ij}=\left(2-\frac{\sol{\beta}_{ij}}{\max(\sol{\beta}_i,\sol{\beta}_j)}\right)^{-\sol{\alpha}_{ij}}\in\bigl[2^{-\sol{\alpha}_{ij}},1\bigr];
\end{equation}
\item[(iii)] in all cases,
\begin{equation}
\label{eq:GammaPareto_etaij}
\eta_{ij}=\left(2-\frac{\sol{\alpha}_{ij}}{\max(\sol{\alpha}_i,\sol{\alpha}_j)}\right)^{-1}\in\left[\tfrac12,1\right],
\end{equation}
with $\eta_{ij}=1$ if and only if $i$ and $j$ belong to the same class, and $\eta_{ij}=\frac12$ if and only if $\A_i\cap\A_j=\varnothing$.
\end{enumerate}
\end{proposition}

\begin{proof}
Let $\tau=(1-u)^{-1}\to\infty$, $s=q_i(u)=\sol{\beta}_i^{-1}(\tau^{1/\sol{\alpha}_i}-1)$ and $t=q_j(u)=\sol{\beta}_j^{-1}(\tau^{1/\sol{\alpha}_j}-1)$, so that $\PP(X_i>s)=\PP(X_j>t)=\tau^{-1}$ and
\begin{align*}
P_\tau&:=\PP(X_i>s,X_j>t)=\e\left[e^{-\Upsilon_\tau}\right],\\
\Upsilon_\tau&:=(\sol{\beta}_i-\sol{\beta}_{ij})sW_i+(\sol{\beta}_j-\sol{\beta}_{ij})tW_j+\sol{\beta}_{ij}\max(sW_i,tW_j).
\end{align*}
Let $\beta_*=\min_{J\in\B_i\cup\B_j}\beta_J>0$. Since $i$ and $j$ are each affected by at least one exponential risk factor, and $\max(a,b)\le a+b$, we have $\tfrac12\beta_*(sW_i+tW_j)\le\beta_*\max(sW_i,tW_j)\le\Upsilon_\tau\le\sol{\beta}_isW_i+\sol{\beta}_jtW_j$. For $c_1,c_2>0$,
\[
\e\left[e^{-c_1sW_i-c_2tW_j}\right]=(1+c_1s)^{-(\sol{\alpha}_i-\sol{\alpha}_{ij})}(1+c_2t)^{-(\sol{\alpha}_j-\sol{\alpha}_{ij})}(1+c_1s+c_2t)^{-\sol{\alpha}_{ij}}.
\]
Assume without loss of generality that $\sol{\alpha}_i\le\sol{\alpha}_j$, so that $s\asymp\tau^{1/\sol{\alpha}_i}$ dominates $t\asymp\tau^{1/\sol{\alpha}_j}$. Both bounds on $P_\tau$ are then of exact order $\tau^{-(\sol{\alpha}_i-\sol{\alpha}_{ij})/\sol{\alpha}_i-(\sol{\alpha}_j-\sol{\alpha}_{ij})/\sol{\alpha}_j-\sol{\alpha}_{ij}/\sol{\alpha}_i}=\tau^{-(2-\sol{\alpha}_{ij}/\sol{\alpha}_j)}$, which proves (iii). The characterization of the endpoints holds because $\sol{\alpha}_{ij}=0$ if and only if $\A_i\cap\A_j=\varnothing$, whereas $\sol{\alpha}_{ij}=\max(\sol{\alpha}_i,\sol{\alpha}_j)=\sol{\alpha}_j$ forces $\A_j\subseteq\A_i$ and $\sol{\alpha}_j\le\sol{\alpha}_i\le\sol{\alpha}_j$, hence $\A_i=\A_j$. For (i), different classes thus give $\eta_{ij}<1$, so that $\chi_{ij}=\lim_{\tau\to\infty}\tau P_\tau=0$. For (ii), $W_i=W_j$ and $\sol{\alpha}_i=\sol{\alpha}_j=\sol{\alpha}_{ij}$, so that $\sol{\beta}_is=\sol{\beta}_jt=\tau^{1/\sol{\alpha}_{ij}}-1=:y$, and
\[
P_\tau=\left(1+y\left(2-\frac{\sol{\beta}_{ij}}{\sol{\beta}_i}-\frac{\sol{\beta}_{ij}}{\sol{\beta}_j}+\frac{\sol{\beta}_{ij}}{\min(\sol{\beta}_i,\sol{\beta}_j)}\right)\right)^{-\sol{\alpha}_{ij}}
=\left(1+y\left(2-\frac{\sol{\beta}_{ij}}{\max(\sol{\beta}_i,\sol{\beta}_j)}\right)\right)^{-\sol{\alpha}_{ij}}.
\]
Since $\tau y^{-\sol{\alpha}_{ij}}\to1$, we obtain \eqref{eq:GammaPareto_Chiij}; the bounds follow from $0\le\sol{\beta}_{ij}\le\min(\sol{\beta}_i,\sol{\beta}_j)$.
\end{proof}

The lower bound $2^{-\sol{\alpha}_{ij}}$ in \eqref{eq:GammaPareto_Chiij} is attained when $\B_i\cap\B_j=\varnothing$, in which case $(X_i,X_j)$ is a bivariate Pareto pair with a common gamma denominator, i.e., the classical model of the second kind. The upper bound $1$ is attained when $\B_i=\B_j$, in which case $X_i=X_j$ almost surely. Strikingly, the coefficient $\eta_{ij}$ in \eqref{eq:GammaPareto_etaij} depends on $\A$ alone. In particular, a pair driven by disjoint gamma factors has $\eta_{ij}=\frac12$, the value under exact independence, even if it shares systemic exponential background risk factors. Exponential risk factors, being light-tailed, shape the body of the joint distribution and the constants in its tail, but not the rate at which joint extremes vanish.

\begin{remark}
\label{rm:GammaPareto_regimes}
Corollary~\ref{cor:GammaPareto_lambda} and Proposition~\ref{prop:GammaPareto_chi} show that the gamma system $\A$ completely determines which tail dependence coefficients are positive, while $\B$ influences only their magnitude once positivity is established. Four regimes arise, summarized in Table~\ref{tab:regimes}. If $\A_i\cap\A_j=\varnothing$, the pair is extremally as weakly dependent as possible. If the factor sets overlap without being nested, the pair is asymptotically independent under both $\lambda$ and $\chi$, but $\eta_{ij}\in(\frac12,1)$ reveals residual extremal dependence. The strictly nested case $\A_j\subsetneq\A_i$ is particularly interesting. Here $\chi_{ij}=0$ while $\lambda_{ij}>0$: the variables are asymptotically independent in the copula sense, yet retain positive common-threshold tail dependence and $\eta_{ij}\in(\frac12,1)$. Such a separation between vanishing classical coefficients and persistent extremal dependence is characteristic of hidden regular variation \citep{Resnick2002}. Finally, if $i$ and $j$ belong to the same class, all coefficients are positive and $\eta_{ij}=1$.
\end{remark}

\begin{table}[H]
\centering
\caption{Extremal dependence regimes of a pair $(X_i,X_j)$, $i\neq j$, in terms of the gamma factor sets $\A_i$ and $\A_j$.}
\label{tab:regimes}
\renewcommand{\arraystretch}{1.3}
\begin{tabular}{lcccc}
\hline
Regime & $\lambda_{ij}$ & $\lambda_{ji}$ & $\chi_{ij}$ & $\eta_{ij}$\\
\hline
$\A_i\cap\A_j=\varnothing$ & $0$ & $0$ & $0$ & $\frac12$\\
overlapping, not nested & $0$ & $0$ & $0$ & $\in(\frac12,1)$\\
$\A_j\subsetneq\A_i$ & $>0$, eq.~\eqref{eq:GammaPareto_Lambdaij} & $0$ & $0$ & $\in(\frac12,1)$\\
$\A_i=\A_j$ (same class) & $>0$, eq.~\eqref{eq:GammaPareto_Lambdaij} & $>0$ & $>0$, eq.~\eqref{eq:GammaPareto_Chiij} & $1$\\
\hline
\end{tabular}
\end{table}

\begin{example}
\label{ex:GammaPareto_BivariateExtremal}
In Example~\ref{ex:GammaPareto_Bivariate}, Cases (1) and (2) have disjoint gamma factor sets, so that $\lambda_{12}=\lambda_{21}=\chi_{12}=0$ and $\eta_{12}=\frac12$; in Case (2) this holds although $\Cov(X_1,X_2)>0$ whenever $\alpha_1,\alpha_2>1$. Cases (3) and (4) are overlapping but not nested, with
\[
\eta_{12}=\left(2-\frac{\alpha_{12}}{\max(\alpha_1,\alpha_2)+\alpha_{12}}\right)^{-1}\in\left(\tfrac12,1\right).
\]
Dropping the idiosyncratic gamma risk factor $G_2$ produces the nested configuration $\A_2\subsetneq\A_1$, for which $\lambda_{12}=\bigl(\sol{\beta}_1/(\sol{\beta}_1+\sol{\beta}_2-\sol{\beta}_{12})\bigr)^{\alpha_{12}}>0$ while $\chi_{12}=0$. Finally, in Case (5), both components belong to the same class, $\eta_{12}=1$, and
\[
\chi_{12}=\left(2-\frac{\beta_{12}}{\max(\beta_1,\beta_2)+\beta_{12}}\right)^{-\alpha_{12}}.
\]
\end{example}

Taken together, the results of this section complete the picture drawn by Theorem~\ref{thm:mainPD}. The sub-mixture representation distinguishes the contributions of the Marshall--Olkin and additive gamma risk factor structures at the level of the survival function, while the covariance decomposition \eqref{eq:GammaPareto_CovarianceDecomposition} separates their contributions and their interaction. The extremal analysis shows that the gamma risk factors determine the presence of asymptotic tail dependence and the decay rates of joint exceedance probabilities, whereas the exponential risk factors affect the associated tail constants.

\section{Conclusions}
\label{sec:conclusions}

This paper revisits multiplicative background risk models with Pareto
margins when Marshall--Olkin exponential and additive gamma risk factor
structures act simultaneously. Our central finding,
Theorem~\ref{thm:mainPD}, is that, under an admissibility condition on
the risk factor systems, the resulting joint survival function admits
a finite sub-mixture representation in terms of Pareto survival
functions. The contribution of each systemic exponential background
risk factor is assigned to the class imposing the largest effective
threshold among those it affects, and the weights are gamma, or
equivalently Dirichlet, probabilities of the corresponding ordering
regions after exponential tilting. Conventionally read by conditioning
on the gamma background risk factors, the multiplicative background
risk model thus acquires a complementary reading directly through
Pareto survival functions and their Dirichlet weights. The sub-mixture
deficit quantifies the dependence between the class orderings and
the exceedances of the auxiliary risk vectors.

The representation also unifies and extends the classical constructions
of \cite{Arnold2015} within a single tractable family. The family is
closed under marginalization, has Pareto margins and tractable
distributions of minima, and admits a complete characterization of
independence and, under a non-degeneracy condition, of its singular
and absolutely continuous parts. Its dependence structure exhibits
a clear division of labour. Both risk factor structures contribute
to association, and the covariance decomposition separates their
contributions and their interaction. In the tails, however, the
gamma risk factors alone determine the presence of asymptotic tail
dependence and the decay rates of joint exceedance probabilities,
whereas the exponential risk factors affect only the associated
tail constants.

Several directions remain open. Statistical inference must accommodate
singular components and threshold-dependent sub-mixture weights,
efficient algorithms are needed for the Dirichlet weights in high
dimensions, and the explicit structure invites applications to risk
measures, capital allocation and reinsurance for collections of risks
exposed to systemic background risk factors.


\section*{Acknowledgements}

This work received financial support from the NORM program
through NSERC grant ALLRP 580632 and Mitacs grant IT33381.

We dedicate this paper to the memory of Professor Ri\v{c}ardas
Zitikis. Edward Furman remembers him with deep affection as
a dear friend and cherishes their discussions on multiplicative
background risk models, a topic to which Ri\v{c}ardas made valued
contributions, alongside his many contributions to quantitative
risk management, actuarial science, and beyond. Nawaf Mohammed
remembers him with gratitude as a member of his doctoral committee.

\section*{Declaration of competing interest}

The authors declare that they have no known competing financial
interests or personal relationships that could have appeared to
influence the work reported in this paper.
\section*{Data availability}

No data was used for the research described in the article.

\newpage

\newpage
\clearpage
\begin{appendices}

\section{Proofs of the results in Section~\ref{sec:GammaPareto}}
\label{app:proofs_sec2}

Throughout this appendix, $\PP_{\GG}$ denotes the law of $\GG=(G_I,I\in\A)$, which is absolutely continuous on $(0,\infty)^{|\A|}$. We use repeatedly the fact that every hyperplane in $\R^{|\A|}$ is $\PP_{\GG}$-null.

\begin{proof}[Proof of Theorem~\ref{thm:mainPD}]
Fix $\xx\in[0,\infty)^n\setminus\{\mathbf{0}\}$, the case $\xx=\mathbf{0}$ being trivial, and abbreviate $\Lambda_I^{\pi}=\Lambda_I^{\pi}(\xx)$, $\Omega^{\pi}=\Omega^{\pi}(\xx)$ and $\widetilde{\Omega}^{\pi}=\widetilde{\Omega}^{\pi}(\xx)$.

\emph{Step 1: conditioning.} Since $\{V_i>x_iW_i\ \forall i\in N\}=\{E_J>\max_{j\in J}x_jW_j\ \forall J\in\B\}$, and the $E_J$ are independent of each other and of $\WW$,
\begin{equation}
\label{eq:app_laplace}
\SF_{\XX}(\xx)=\e\left[\exp\left\{-\sum_{J\in\B}\beta_J\max_{j\in J}\left(x_jW_j\right)\right\}\right].
\end{equation}
By \eqref{eq:class_denominator}, $\max_{j\in J}(x_jW_j)=\max_{j\in J}(x_j)\,W^{(k_J)}$ for $J\in\B_{-}$. For $J\in\B_{+}$, admissibility gives $\max_{j\in J}(x_jW_j)=\max_{k\in M_J}x_k^{*}W^{(k)}$, which equals $\max_{k\in M_J\cap M_+}x_k^{*}W^{(k)}$ if $J\in\B_{+}(\xx)$ and vanishes otherwise.

\emph{Step 2: latent orderings.} For $k\in M_+$, let $\widetilde{\Psi}_k(\gg)=x_k^{*}\sum_{I\in\AA_k}g_I$, so that $\widetilde{\Psi}_k(\GG)=x_k^{*}W^{(k)}$ and $\widetilde{\Omega}^{\pi}=\{\widetilde{\Psi}_{\pi(1)}\ge\cdots\ge\widetilde{\Psi}_{\pi(m_+)}\}$. For $k\neq\ell$ in $M_+$, the linear form $\widetilde{\Psi}_k-\widetilde{\Psi}_\ell$ has a non-zero coefficient at every $I$ in the non-empty symmetric difference of $\AA_k$ and $\AA_\ell$, so that $\{\widetilde{\Psi}_k=\widetilde{\Psi}_\ell\}$ is a hyperplane. Consequently, the cones $\widetilde{\Omega}^{\pi}$, $\pi\in\SS_{M_+}$, cover $[0,\infty)^{|\A|}$ and their pairwise intersections are $\PP_{\GG}$-null. On $\{\GG\in\widetilde{\Omega}^{\pi}\}$, the maximum over $M_J\cap M_+$ in Step 1 is attained at the leading class $k_J^{\pi}$ for every $J\in\B_{+}(\xx)$. Using $W^{(k)}=\sum_{I\in\A:\,k\in M_I}G_I$, which follows from \eqref{eq:incidence}, and interchanging the order of summation, we obtain on $\{\GG\in\widetilde{\Omega}^{\pi}\}$
\[
\sum_{J\in\B}\beta_J\max_{j\in J}\left(x_jW_j\right)
=\sum_{J\in\B_{-}}\beta_J\max_{j\in J}\left(x_j\right)W^{(k_J)}+\sum_{J\in\B_{+}(\xx)}\beta_J\,x^{*}_{k_J^{\pi}}W^{(k_J^{\pi})}
=\sum_{I\in\A}\Lambda_I^{\pi}\,G_I.
\]
Hence
\begin{equation}
\label{eq:app_partition}
\SF_{\XX}(\xx)=\sum_{\pi\in\SS_{M_+}}\e\left[\exp\left\{-\sum_{I\in\A}\Lambda_I^{\pi}G_I\right\}\mathds{1}_{\{\GG\in\widetilde{\Omega}^{\pi}\}}\right].
\end{equation}

\emph{Step 3: exponential tilting.} For $\cc=(c_I,I\in\A)\in[0,\infty)^{|\A|}$ and any Borel set $B\subseteq[0,\infty)^{|\A|}$,
\[
\e\left[e^{-\sum_{I\in\A}c_IG_I}\mathds{1}_{\{\GG\in B\}}\right]
=\prod_{I\in\A}(1+c_I)^{-\alpha_I}\,\PP\left(\left(\frac{G_I}{1+c_I}\right)_{I\in\A}\in B\right),
\]
because $g^{\alpha_I-1}e^{-(1+c_I)g}/\Gamma(\alpha_I)$ equals $(1+c_I)^{-\alpha_I}$ times the $\mathrm{Gamma}(\alpha_I,1+c_I)$ density, and $\mathrm{Gamma}(\alpha_I,1+c_I)$ is the law of $G_I/(1+c_I)$. Applying this identity with $\cc=(\Lambda_I^{\pi})_{I\in\A}$ and $B=\widetilde{\Omega}^{\pi}$, and observing that $(G_I/(1+\Lambda_I^{\pi}))_{I\in\A}\in\widetilde{\Omega}^{\pi}$ if and only if $\GG\in\Omega^{\pi}$ by \eqref{eq:Omega}, we obtain \eqref{eq:GammaPareto} from \eqref{eq:app_partition}.

\emph{Step 4: the sub-mixture inequality.} For $\pi\in\SS_{M_+}$ and $k\in M_+$, define the tilted scores
\[
\Psi_k^{\pi}(\gg)=x_k^{*}\sum_{I\in\AA_k}\frac{g_I}{1+\Lambda_I^{\pi}}=\sum_{I\in\A}\frac{x_k^{*}\,\mathds{1}_{\{k\in M_I\}}}{1+\Lambda_I^{\pi}}\,g_I,
\]
so that $\Omega^{\pi}=\{\Psi_{\pi(1)}^{\pi}\ge\cdots\ge\Psi_{\pi(m_+)}^{\pi}\}$, and let $\Omega^{\pi}_{\circ}=\{\gg\in(0,\infty)^{|\A|}:\Psi_{\pi(1)}^{\pi}>\cdots>\Psi_{\pi(m_+)}^{\pi}\}$. As in Step 2, each $\Psi^{\pi}_k-\Psi^{\pi}_\ell$, $k\neq\ell$, is a non-zero linear form, whence $\PP_{\GG}(\Omega^{\pi}\setminus\Omega^{\pi}_{\circ})=0$. It therefore suffices to show that $\Omega^{\pi_1}_{\circ}\cap\Omega^{\pi_2}_{\circ}=\varnothing$ for $\pi_1\neq\pi_2$, since then
\[
\sum_{\pi\in\SS_{M_+}}\PP\left(\GG\in\Omega^{\pi}\right)=\sum_{\pi\in\SS_{M_+}}\PP\left(\GG\in\Omega^{\pi}_{\circ}\right)=\PP\Bigl(\GG\in\bigcup_{\pi\in\SS_{M_+}}\Omega^{\pi}_{\circ}\Bigr)\le1.
\]
Fix $\pi_1\neq\pi_2$ and call $(k,\ell)$, $k\neq\ell\in M_+$, an \emph{inversion pair} if $k$ precedes $\ell$ under $\pi_1$ and $\ell$ precedes $k$ under $\pi_2$. Define
\[
\rho_{k\ell}=\sum_{\substack{J\in\B_{+}(\xx)\\ k_J^{\pi_1}=k,\ k_J^{\pi_2}=\ell}}\beta_J\ \ge0,
\qquad
\mathcal{L}=\{(k,\ell):k\neq\ell,\ \rho_{k\ell}>0\}.
\]
Every $(k,\ell)\in\mathcal{L}$ is an inversion pair: some $J$ with $k,\ell\in M_J\cap M_+$ has leader $k$ under $\pi_1$ and leader $\ell$ under $\pi_2$. With $\delta_I^{k\ell}=x_k^{*}\mathds{1}_{\{k\in M_I\}}-x_\ell^{*}\mathds{1}_{\{\ell\in M_I\}}$, the $\pi$-independent contributions of $\B_{-}$ cancel, and grouping the exponential risk factors of $\B_{+}(\xx)$ according to their pair of leaders gives
\begin{equation}
\tag{$\star$}
\Lambda_I^{\pi_1}-\Lambda_I^{\pi_2}
=\sum_{J\in\B_{+}(\xx)}\beta_J\left[x^{*}_{k_J^{\pi_1}}\mathds{1}_{\{k_J^{\pi_1}\in M_I\}}-x^{*}_{k_J^{\pi_2}}\mathds{1}_{\{k_J^{\pi_2}\in M_I\}}\right]
=\sum_{(k,\ell)\in\mathcal{L}}\rho_{k\ell}\,\delta_I^{k\ell},
\qquad I\in\A,
\end{equation}
since exponential risk factors with equal leaders contribute zero. Suppose, for a contradiction, that $\gg\in\Omega^{\pi_1}_{\circ}\cap\Omega^{\pi_2}_{\circ}$. For every inversion pair $(k,\ell)$, transitivity of the strict chains yields $\Psi_k^{\pi_1}(\gg)>\Psi_\ell^{\pi_1}(\gg)$ and $\Psi_\ell^{\pi_2}(\gg)>\Psi_k^{\pi_2}(\gg)$. Adding these inequalities gives
\begin{equation}
\tag{$\star\star$}
\sum_{I\in\A}g_I\,\delta_I^{k\ell}\,\vartheta_I>0,
\qquad
\vartheta_I:=\frac{1}{1+\Lambda_I^{\pi_1}}-\frac{1}{1+\Lambda_I^{\pi_2}}.
\end{equation}

\emph{Case $\mathcal{L}=\varnothing$.} By $(\star)$, $\Lambda^{\pi_1}_I=\Lambda^{\pi_2}_I$ for all $I$, hence $\Psi_k^{\pi_1}=\Psi_k^{\pi_2}=:\Psi_k$ for all $k$. Since $\pi_1\neq\pi_2$, some inversion pair $(k,\ell)$ exists, and then $\Psi_k(\gg)>\Psi_\ell(\gg)>\Psi_k(\gg)$, which is absurd.

\emph{Case $\mathcal{L}\neq\varnothing$.} Multiplying $(\star\star)$ by $\rho_{k\ell}>0$, summing over $(k,\ell)\in\mathcal{L}$ and using $(\star)$, we find
\[
0<\sum_{(k,\ell)\in\mathcal{L}}\rho_{k\ell}\sum_{I\in\A}g_I\,\delta_I^{k\ell}\,\vartheta_I
=\sum_{I\in\A}g_I\,\vartheta_I\left(\Lambda_I^{\pi_1}-\Lambda_I^{\pi_2}\right)
=-\sum_{I\in\A}g_I\,\frac{\left(\Lambda_I^{\pi_1}-\Lambda_I^{\pi_2}\right)^{2}}{(1+\Lambda_I^{\pi_1})(1+\Lambda_I^{\pi_2})}\le0,
\]
again a contradiction. Hence $\Omega^{\pi_1}_{\circ}\cap\Omega^{\pi_2}_{\circ}=\varnothing$, which establishes \eqref{eq:GammaPareto_weights_lessthanone}.
\end{proof}

\begin{proof}[Proof of Proposition~\ref{prop:GammaPareto_singular}]
Since $L\subseteq N_k$, all $X_i$, $i\in L$, share the denominator $W^{(k)}$, so that $\Theta_k^L=\{V_i=V_j\ \forall i,j\in L\}$. Let $\BB_{L}=\bigcup_{i\in L}\B_i=\{J\in\B:J\cap L\neq\varnothing\}$, the set of labels of exponential risk factors affecting $L$ (cf.\ $\BB_k$ in Proposition~\ref{prop:GammaPareto_ACvsSingular}), let $T=\min_{J\in\BB_{L}}E_J=\min_{i\in L}V_i$, and let $J^{*}$ be the almost surely unique minimizer. We claim that $\Theta_k^L=\{L\subseteq J^{*}\}$ almost surely. If $L\subseteq J^{*}$, then $J^{*}\in\B_i$ and $E_{J^{*}}=T\le E_J$ for all $J\in\B_i\subseteq\BB_{L}$, so that $V_i=E_{J^{*}}$ for every $i\in L$. Conversely, on $\Theta_k^L$ all $V_i$, $i\in L$, equal $T$; each $V_i$ is attained by a unique $J_i\in\B_i$, and since the $E_J$ are almost surely distinct, all $J_i$ coincide with $J^{*}$, whence $L\subseteq J^{*}$. The exponential race gives $\PP(J^{*}=J)=\beta_J/\sol{\beta}^{L}$ for $J\in\BB_{L}$; summing over $\{J\in\B:L\subseteq J\}$, which is non-empty by assumption, yields the stated probability. Moreover, for $t>0$ and $J\in\BB_{L}$,
\[
\PP\left(T>t,\,J^{*}=J\right)=\int_t^{\infty}\beta_J\,e^{-\sol{\beta}^{L}u}\,\d u=\frac{\beta_J}{\sol{\beta}^{L}}\,e^{-\sol{\beta}^{L}t}=\PP(J^{*}=J)\,\PP(T>t),
\]
so that $T$ is independent of $J^{*}$, and hence of the $\sigma(J^{*})$-measurable event $\Theta_k^L$. Since $W^{(k)}\sim\mathrm{Gamma}(\sol{\alpha}^{(k)},1)$ is independent of $(T,J^{*})$, the minimum $T^{L}=T/W^{(k)}$ is independent of $\Theta_k^L$, and $X_i=T^{L}$ on $\Theta_k^L$ for every $i\in L$. Therefore,
\[
\PP\left(X_i>x\,\middle|\,\Theta_k^L\right)=\PP\left(T^{L}>x\right)=\e\left[e^{-\sol{\beta}^{L}xW^{(k)}}\right]=\left(1+\sol{\beta}^{L}x\right)^{-\sol{\alpha}^{(k)}},\qquad x\ge0,
\]
which also proves the identity $X_i\mid\Theta_k^L\overset{d}{=}T^{L}$ noted after the proposition.
\end{proof}

For the proof of Proposition~\ref{prop:GammaPareto_ACvsSingular}, recall that the $E_J$ are almost surely distinct. For each $i\in N$, we say that the exponential risk factor indexed by $c(i)\in\B_i$ with $E_{c(i)}=V_i$ \emph{claims} $i$, and we call the random map $c:N\to\B$ the \emph{claiming pattern}. Since within-class ties of $\XX$ and $\VV$ coincide, we have $\{\XX\notin\Theta\}=\{V_i\neq V_j$ for all $i\neq j$ in the same class$\}$.

\begin{lemma}
\label{lem:claims}
On $\{\XX\notin\Theta\}$: (a) $c$ is injective on every class; (b) every class contains at most one index claimed by an exponential risk factor in $\B_{+}$; (c) the indices claimed by exponential risk factors in $\B_{-}$ are claimed by pairwise distinct factors.
\end{lemma}

\begin{proof}
(a) If $c(i)=c(j)$, then $V_i=V_j$. (b) Let $i\neq i'$ in $N_k$ be claimed by $J,J'\in\B_{+}$. By admissibility, $N_k\subseteq J\cap J'$, so that $V_{i'}\le E_J=V_i$ and $V_i\le E_{J'}=V_{i'}$, i.e., $V_i=V_{i'}$, which is impossible. (c) An exponential risk factor in $\B_{-}$ is confined to a single class, and (a) applies.
\end{proof}

\begin{lemma}
\label{lem:ac_offdiag}
If \eqref{eq:ND} holds, then the measure $\PP(\XX\in\cdot\,,\,\XX\notin\Theta)$ is absolutely continuous.
\end{lemma}

\begin{proof}
Let $Z\subseteq\R^n$ be Lebesgue-null. As the claiming pattern takes finitely many values, it suffices to show that $\PP(\XX\in Z,\,C)=0$, where $C=\{c=\gamma\}\cap\{\XX\notin\Theta\}$ for a fixed map $\gamma$ with $\PP(C)>0$; by Lemma~\ref{lem:claims}, $\gamma$ inherits properties (a)--(c). Denote by $k(i)$ the class of $i$, let $N_{\pm}=\{i:\gamma(i)\in\B_{\pm}\}$, and define $\YY=(Y_i,i\in N)$ by $Y_i=E_{\gamma(i)}/W^{(k(i))}$, so that $\XX=\YY$ on $C$. By property (b), the classes $k(i)$, $i\in N_+$, are pairwise distinct elements of $M_{\B_{+}}$. Hence, conditionally on $\mathcal{E}=\sigma(E_J,J\in\B)$, the vector $\YY^{N_+}=(Y_i,i\in N_+)$ is the image of $(W^{(k(i))})_{i\in N_+}$ under a coordinatewise diffeomorphism of $(0,\infty)^{|N_+|}$. The latter vector is absolutely continuous by \eqref{eq:ND} and independent of $\mathcal{E}$, so that $\PP(\YY^{N_+}\in Z')=0$ for every Lebesgue-null $Z'\subseteq\R^{|N_+|}$. Next, let $D_-=\{\gamma(i):i\in N_-\}$ and $\mathcal{G}=\sigma(\WW,\,E_J,\,J\in\B\setminus D_-)$. The vector $\YY^{N_+}$ is $\mathcal{G}$-measurable, whereas, by property (c), $(E_{\gamma(i)})_{i\in N_-}$ consists of distinct independent exponentials that are independent of $\mathcal{G}$. Consequently, given $\mathcal{G}$, the vector $\YY^{N_-}=(Y_i,i\in N_-)$ is a linear bijective image of an absolutely continuous vector. Writing $Z_u=\{y:(u,y)\in Z\}$ for the sections of $Z$, the set $Z'=\{u:Z_u\text{ is not Lebesgue-null}\}$ is Lebesgue-null by Fubini's theorem, and $\PP(\YY\in Z\mid\mathcal{G})=0$ on $\{\YY^{N_+}\notin Z'\}$. Therefore
\[
\PP(\XX\in Z,\,C)\le\PP(\YY\in Z)\le\PP\left(\YY^{N_+}\in Z'\right)+\e\left[\mathds{1}_{\{\YY^{N_+}\notin Z'\}}\PP\left(\YY\in Z\mid\mathcal{G}\right)\right]=0.
\]
The degenerate cases $N_+=\varnothing$ and $N_-=\varnothing$ are covered by the same argument.
\end{proof}

\begin{proof}[Proof of Proposition~\ref{prop:GammaPareto_ACvsSingular}]
Since $\Theta$ is a finite union of hyperplanes, $\PP_{\ac}(\Theta)=0$. Under \eqref{eq:ND}, Lemma~\ref{lem:ac_offdiag} and the uniqueness of the Lebesgue decomposition yield $\PP_{\co}=\PP_{\XX}(\,\cdot\cap\Theta)$ and $\PP_{\ac}=\PP_{\XX}(\,\cdot\setminus\Theta)$.

\emph{(i) and (ii).} If $i\neq j\in N_k$ and $J\in\B_i\cap\B_j$, then $\{X_i=X_j\}\supseteq\{E_J<E_{J'}\ \forall J'\in(\B_i\cup\B_j)\setminus\{J\}\}$, so that $\PP_{\XX}(\Theta)>0$. Since $\Theta$ is Lebesgue-null, $\PP_{\ac}(\Theta)=0$ and therefore $\PP_{\co}(\Theta)=\PP_{\XX}(\Theta)>0$; hence $\PP_{\co}\neq0$, without recourse to \eqref{eq:ND}. Conversely, if $\B_i\cap\B_j=\varnothing$ for all $i\neq j$ in the same class, then $V_i$ and $V_j$ are functions of disjoint families of independent continuous variables, so that $\PP(X_i=X_j)=\PP(V_i=V_j)=0$ and $\PP_{\XX}(\Theta)=0$; under \eqref{eq:ND}, $\PP_{\co}=\PP_{\XX}(\,\cdot\cap\Theta)=0$. This proves (i), and (ii) is its negation.

\emph{(iii) and (iv).} Suppose first that $\PP_{\ac}\neq0$. Then $\PP(\XX\notin\Theta)\ge\PP_{\ac}(\Theta^{c})>0$, and there is an outcome with distinct $E_J$ and no within-class tie. Fix a class $N_k$ with $|N_k|\ge2$ and list its elements $i_1,\dots,i_{|N_k|}$ in increasing order of the arrival times $V_{i}=E_{c(i)}$, which are distinct by Lemma~\ref{lem:claims}(a). When $J_t:=c(i_t)$ arrives, the indices of $N_k$ not yet claimed are exactly $U_t$; since $J_t$ claims every such index it contains, and it claims $i_t$ only, $J_t\cap U_t=\{i_t\}$. Thus $N_k$ admits a separating order; this direction does not use \eqref{eq:ND}.

Conversely, suppose that every class $N_k$ with $|N_k|\ge2$ admits a separating order with witnesses $J_1^{(k)},\dots,J_{|N_k|-1}^{(k)}$. For $t<|N_k|$ we have $|U_t|\ge2$, and an exponential risk factor in $\B_{+}$ affecting $N_k$ affects the entire class; hence every witness belongs to $\B_{-}$ and is contained in $N_k$. Witnesses of distinct classes are therefore distinct. Within a class they are distinct as well: for $s<t$, we have $i_t\in J^{(k)}_t$ but $i_t\notin J^{(k)}_s$, because $J^{(k)}_{s}\cap U_{s}=\{i_s\}$ and $i_t\in U_s\setminus\{i_s\}$. Let $F$ be the event that all exponential risk factors indexed by the witnesses arrive before all other exponential risk factors and that, within each class, the corresponding exponential risk factors arrive in the order $E_{J_1^{(k)}}<\cdots<E_{J_{|N_k|-1}^{(k)}}$. The event $F$ contains the event that the finitely many independent exponentials $E_J$ realize a fixed total order compatible with these constraints, so that $\PP(F)>0$. On $F$, fix a class $N_k$; exponential risk factors of other witness chains do not affect $N_k$. We show by induction that $J_t^{(k)}$ claims exactly $i_t$: when it arrives, the unclaimed indices of $N_k$ are $U_t$, and $J_t^{(k)}\cap U_t=\{i_t\}$. After the last witness, only $i_{|N_k|}$ remains unclaimed and is subsequently claimed alone. Hence $F\subseteq\{\XX\notin\Theta\}$, and, under \eqref{eq:ND}, $\PP_{\ac}(\R^n)=\PP(\XX\notin\Theta)\ge\PP(F)>0$. This proves (iii), and (iv) is its negation.
\end{proof}

\section{Illustrative examples}
\label{app:examples}

This appendix works through two examples. The first applies the results of Sections~\ref{sec:GammaPareto} and~\ref{sec:Dependence_Measures} to a trivariate risk vector that is covered neither by \eqref{eq:6.2.8arnold} and \eqref{eq:6.2.3arnold} nor by the motivating example \eqref{eq:motiv_model}. The second relates the sub-mixture deficit of the motivating example to the extremal dependence results of Section~\ref{subsec:extremal}.

\begin{example}[Two regions with regional and global risk factors]
\label{ex:B_regional}
Let $n=3$, and think of the first two risk components as belonging to one region and the third to another. The gamma risk factor system $\A=\{\{1,2\},\{3\},N\}$ consists of a factor for each region and a global factor, and the exponential risk factor system $\B=\{\{1\},\{2\},\{3\},\{1,2\},N\}$ consists of idiosyncratic factors, a regional systemic factor and a global systemic factor. As in Example~\ref{ex:GammaPareto_Bivariate}, we drop braces from labels and write $G_{12}$, $G_3$, $G_N$ for the gamma risk factors, with shapes $\alpha_{12}$, $\alpha_3$, $\alpha_N$, and $\beta_1$, $\beta_2$, $\beta_3$, $\beta_{12}$, $\beta_N$ for the rates of the exponential risk factors.

\medskip
\noindent\emph{Factor-selection framework.} Since $\A_1=\A_2=\{\{1,2\},N\}$ and $\A_3=\{\{3\},N\}$, there are $m=2$ classes, $N_1=\{1,2\}$ and $N_2=\{3\}$, with class denominators $W^{(1)}=G_{12}+G_N$ and $W^{(2)}=G_3+G_N$. Every element of $\B$ is admissible: $\{1\}$, $\{2\}$, $\{1,2\}$ and $\{3\}$ belong to $\B_{-}$, whereas $N=N_1\cup N_2$ belongs to $\B_{+}$. As $|M_{\B_{+}}|=2$, condition \eqref{eq:ND} holds. The model is covered by none of \eqref{eq:6.2.8arnold}, \eqref{eq:6.2.3arnold} and \eqref{eq:motiv_model}: the global exponential risk factor affects two classes with different gamma denominators, while the regional one affects two components of the same class.

\medskip
\noindent\emph{Sub-mixture representation.} Let $\xx\in(0,\infty)^3$, write $x_1^{*}=\max(x_1,x_2)$ and $x_2^{*}=x_3$ for the class thresholds, and let $\ell=\beta_1x_1+\beta_2x_2+\beta_{12}x_1^{*}$ and $r=\beta_3x_3$ collect the contributions of the exponential risk factors confined to the two classes. The only element of $\B_{+}$ is $N$, whose leading class is $\pi(1)$. By \eqref{eq:Lambda_MP},
\begin{align*}
\pi=(1,2):&\quad \Lambda_{12}^{\pi}=\ell+\beta_Nx_1^{*},\qquad \Lambda_{3}^{\pi}=r,\qquad\quad\ \ \Lambda_N^{\pi}=\ell+r+\beta_Nx_1^{*},\\
\pi=(2,1):&\quad \Lambda_{12}^{\pi}=\ell,\qquad\qquad\ \ \ \Lambda_{3}^{\pi}=r+\beta_Nx_3,\quad \Lambda_N^{\pi}=\ell+r+\beta_Nx_3,
\end{align*}
and Theorem~\ref{thm:mainPD} gives
\begin{align*}
\SF_{\XX}(\xx)&=w^{(1,2)}(\xx)\,\bigl(1+\ell+\beta_Nx_1^{*}\bigr)^{-\alpha_{12}}(1+r)^{-\alpha_3}\bigl(1+\ell+r+\beta_Nx_1^{*}\bigr)^{-\alpha_N}\\
&\quad+w^{(2,1)}(\xx)\,(1+\ell)^{-\alpha_{12}}\bigl(1+r+\beta_Nx_3\bigr)^{-\alpha_3}\bigl(1+\ell+r+\beta_Nx_3\bigr)^{-\alpha_N},
\end{align*}
where $w^{\pi}(\xx)=\PP(\GG\in\Omega^{\pi}(\xx))$, $\Omega^{(1,2)}(\xx)$ is the half-space of points $\gg=(g_{12},g_3,g_N)$ satisfying
\[
x_1^{*}\left(\frac{g_{12}}{1+\Lambda_{12}^{\pi}}+\frac{g_N}{1+\Lambda_N^{\pi}}\right)\ge x_3\left(\frac{g_3}{1+\Lambda_3^{\pi}}+\frac{g_N}{1+\Lambda_N^{\pi}}\right)
\]
with the values of $\Lambda^{\pi}$ for $\pi=(1,2)$, and $\Omega^{(2,1)}(\xx)$ is defined by the reverse inequality with the values for $\pi=(2,1)$. In the language of Remark~\ref{rm:GammaPareto_interpretation}, the first component is the survival function of the auxiliary risk vector in which the global exponential risk factor is merged with the regional one, so that the first class faces the rate $\beta_{12}+\beta_N$ and the third component only its idiosyncratic factor; the second component assigns the global factor to the third component instead. By \eqref{eq:Dirichlet_general}, each weight is the probability that $\RR=(R_{12},R_3,R_N)\sim\mathrm{Dirichlet}(\alpha_{12},\alpha_3,\alpha_N)$ falls in a triangle cut from the simplex by a single line.

\medskip
\noindent\emph{Weights along the diagonal.} For $\xx=t\one$, $t>0$, the class thresholds coincide, so the coefficient $(x_1^{*}-x_2^{*})/(1+\Lambda_N^{\pi})$ of $g_N$ in the inequality defining $\Omega^{\pi}(t\one)$ vanishes. The global gamma factor scales both classes alike and therefore cannot decide which class absorbs the global exponential factor; only $B:=G_{12}/(G_{12}+G_3)\sim\mathrm{Beta}(\alpha_{12},\alpha_3)$ matters. Writing $I_u(a,b)$ for the regularized incomplete beta function and
\[
u^{\pi}(t)=\frac{1+\Lambda_{12}^{\pi}(t\one)}{2+\Lambda_{12}^{\pi}(t\one)+\Lambda_3^{\pi}(t\one)},\qquad \pi\in\{(1,2),(2,1)\},
\]
we obtain $w^{(1,2)}(t\one)=\PP\bigl(B\ge u^{(1,2)}(t)\bigr)$ and $w^{(2,1)}(t\one)=\PP\bigl(B\le u^{(2,1)}(t)\bigr)$. Since $u^{(2,1)}(t)<u^{(1,2)}(t)$, the sub-mixture deficit has the explicit form
\begin{equation}
\label{eq:B_regional_deficit}
1-w^{(1,2)}(t\one)-w^{(2,1)}(t\one)=\PP\bigl(u^{(2,1)}(t)<B<u^{(1,2)}(t)\bigr)=I_{u^{(1,2)}(t)}(\alpha_{12},\alpha_3)-I_{u^{(2,1)}(t)}(\alpha_{12},\alpha_3).
\end{equation}
It is the probability that the regional gamma factors are too balanced for either class to absorb the global exponential factor after exponential tilting. The interval $(u^{(2,1)}(t),u^{(1,2)}(t))$ has length $\beta_Nt/\{2+t(\beta_1+\beta_2+\beta_3+\beta_{12}+\beta_N)\}$: it vanishes with $\beta_N$, in which case $\B_{+}=\varnothing$ and the weights sum to one, and its length increases with $t$. As $t\downarrow0$, both $u^{\pi}(t)$ tend to $\frac12$ and the deficit vanishes. As $t\to\infty$, Proposition~\ref{prop:GammaPareto_RV} applies with $\B_i\cap\B_{-}\neq\varnothing$ for every $i$, and the weights converge to the strictly positive limits $\PP(\GG\in\Omega_\infty^{\pi}(\one))$, so that the deficit converges to a limit in $(0,1)$.

\medskip
\noindent\emph{Numerical illustration.} Let $\alpha_{12}=\alpha_3=2$, $\alpha_N=1$, $\beta_1=\beta_3=1$ and $\beta_2=\beta_{12}=\beta_N=\frac12$, so that $\sol{\alpha}_1=\sol{\alpha}_2=\sol{\alpha}_3=3$, $\sol{\beta}_1=2$ and $\sol{\beta}_2=\sol{\beta}_3=\frac32$. Here $I_u(2,2)=u^2(3-2u)$, and Table~\ref{tab:B_regional} reports the weights, the deficit \eqref{eq:B_regional_deficit}, the joint survival probability and, for comparison, the product of the marginal survival probabilities. In the limit $t\to\infty$, the weights equal $\frac{68}{343}$ and $\frac{208}{343}$, the deficit equals $\frac{67}{343}$, and $t^{5}\,\SF_{\XX}(t\one)\to H(\one)\approx0.0283$, where the index $5=\alpha_{12}+\alpha_3+\alpha_N$ is that of Proposition~\ref{prop:GammaPareto_RV}. The joint survival probabilities agree with Monte Carlo estimates based on $4\times10^6$ simulations of \eqref{eq:main_rep}, \eqref{eq:WW_Add_rep} and \eqref{eq:VV_MO_rep} to within simulation error. The ratio of the last two columns grows from about $2.7$ to about $1300$, a strong manifestation of the PUOD property of Corollary~\ref{cor:GammaPareto_PUODPLODCov}.

\begin{table}[H]
\centering
\caption{Sub-mixture weights, deficit and joint survival probabilities along the diagonal in Example~\ref{ex:B_regional}.}
\label{tab:B_regional}
\renewcommand{\arraystretch}{1.25}
\begin{tabular}{cccccc}
\hline
$t$ & $w^{(1,2)}(t\one)$ & $w^{(2,1)}(t\one)$ & deficit & $\SF_{\XX}(t\one)$ & $\prod_{i=1}^3\SF_{X_i}(t)$\\
\hline
$1/4$ & $0.4027$ & $0.5326$ & $0.0647$ & $1.188\times10^{-1}$ & $4.384\times10^{-2}$\\
$1$ & $0.3005$ & $0.5680$ & $0.1315$ & $3.607\times10^{-3}$ & $1.517\times10^{-4}$\\
$4$ & $0.2319$ & $0.5933$ & $0.1748$ & $1.508\times10^{-5}$ & $1.166\times10^{-8}$\\
$\infty$ & $0.1983$ & $0.6064$ & $0.1953$ & -- & --\\
\hline
\end{tabular}
\end{table}

\medskip
\noindent\emph{Margins, ties and dependence measures.} By Proposition~\ref{prop:GammaPareto_marginalization}, $(X_1,X_2)$ is a single-class model with merged gamma shape $\alpha_{12}+\alpha_N$ and exponential rates $\beta_1$, $\beta_2$ and $\beta_{12}+\beta_N$, so that
\[
\SF_{(X_1,X_2)}(x_1,x_2)=\bigl(1+\beta_1x_1+\beta_2x_2+(\beta_{12}+\beta_N)\max(x_1,x_2)\bigr)^{-(\alpha_{12}+\alpha_N)},
\]
the bivariate case of \eqref{eq:6.2.3arnold}, whereas $(X_1,X_3)$ is the model of Case~(4) of Example~\ref{ex:GammaPareto_Bivariate}, in which the regional factors $G_{12}$ and $G_3$ play the roles of the idiosyncratic gamma risk factors, $G_N$ that of the systemic one, and the exponential risk factors have rates $\beta_1+\beta_{12}$, $\beta_3$ and $\beta_N$. A three-dimensional sub-mixture therefore has margins of both classical types. By Proposition~\ref{prop:GammaPareto_singular}, $\PP(X_1=X_2)=(\beta_{12}+\beta_N)/(\beta_1+\beta_2+\beta_{12}+\beta_N)$, which equals $0.4$ for the parameters above, and, as $\{1\}$ is a witness for the enumeration $(1,2)$ of $N_1$, Proposition~\ref{prop:GammaPareto_ACvsSingular} shows that the law of $\XX$ has both a singular part, carried by $\{x_1=x_2\}$, and an absolutely continuous part. For the parameters above, Corollary~\ref{cor:GammaPareto_Covariances} gives $\mathrm{Corr}(X_1,X_2)=\frac35$ and, since ${}_3F_2(1,1,1;3,3;1)=4\sum_{k\ge1}k^{-2}(k+1)^{-2}=\frac43\pi^2-12$,
\[
\C_W=\Cov\!\left(\frac{1}{W_1},\frac{1}{W_3}\right)=\frac{\pi^2}{3}-\frac{13}{4},\qquad \mathrm{Corr}(X_1,X_3)=\frac{14}{9}\left(\frac{\pi^2}{3}-\frac{13}{4}\right)+\frac{1}{18}\approx0.118.
\]
The extremal measures follow from Table~\ref{tab:regimes}. Components $1$ and $2$ share a class, so that $\eta_{12}=1$, $\chi_{12}=(\frac32)^{-3}=\frac{8}{27}$, $\lambda_{12}=(\frac45)^3=0.512$ and $\lambda_{21}=(\frac35)^3=0.216$; the asymmetry of the common-threshold coefficients reflects the different marginal rates $\sol{\beta}_1\neq\sol{\beta}_2$. Components $1$ and $3$ have overlapping but not nested gamma factor sets, so that $\lambda_{13}=\lambda_{31}=\chi_{13}=0$, yet $\eta_{13}=(2-\frac13)^{-1}=\frac35$ reveals residual extremal dependence, generated by the global gamma factor alone. A modest correlation of $0.118$ thus coexists with a coefficient of tail dependence well above the independence value $\frac12$.
\end{example}

\begin{example}[Sub-mixture deficit and tail dependence]
\label{ex:B_deficit}
Consider the motivating example \eqref{eq:motiv_model} with equal shapes $\alpha_1=\cdots=\alpha_n=\alpha$. Then $\XX$ is exchangeable, $\SF_{X_i}(x)=(1+\beta_Nx)^{-\alpha}$ for all $i\in N$, and \eqref{eq:sub-m} evaluated at $\xx=x\one$, $x>0$, yields
\[
\sum_{i\in N}p_i(x\one)=\frac{\SF_{\XX}(x\one)}{\SF_{X_1}(x)}=\PP\bigl(X_2>x,\ldots,X_n>x\,\big|\,X_1>x\bigr).
\]
The total weight is thus the conditional probability that a marginal exceedance is accompanied by a joint exceedance, and the deficit is the conditional probability that it is not. As $x\downarrow0$, the total weight tends to one. As $x\to\infty$, Proposition~\ref{prop:GammaPareto_RV} applies with $\A_+(\one)=\A$ and index $n\alpha$, and, writing $F(y)=y^{\alpha}/\Gamma(\alpha+1)$, the limit function equals
\[
H(\one)=\int_{(0,\infty)^n}e^{-\beta_N\max_{i\in N}u_i}\prod_{i\in N}\frac{u_i^{\alpha-1}}{\Gamma(\alpha)}\,\d\uu
=\int_0^\infty e^{-\beta_Ny}\,\d\bigl(F(y)^n\bigr)
=\frac{n\,\Gamma(n\alpha)}{\Gamma(\alpha)\,\Gamma(\alpha+1)^{n-1}\,\beta_N^{n\alpha}}.
\]
Consequently,
\[
\sum_{i\in N}p_i(x\one)\sim C_{n,\alpha}\,(\beta_Nx)^{-(n-1)\alpha},\qquad C_{n,\alpha}=\frac{n\,\Gamma(n\alpha)}{\Gamma(\alpha)\,\Gamma(\alpha+1)^{n-1}},\qquad x\to\infty,
\]
so that the deficit converges to one, in agreement with $\lambda_{ij}=0$ for all $i\neq j$ by Corollary~\ref{cor:GammaPareto_lambda}(i), at a polynomial rate governed by the gamma shapes alone. For $\alpha=1$, the maximum of $n$ independent unit exponentials has Laplace transform $\prod_{k=1}^n k/(k+s)$, and the total weight is available in closed form,
\[
\sum_{i\in N}p_i(x\one)=\prod_{k=2}^{n}\frac{k}{k+\beta_Nx},
\]
which reduces to $2/(2+\beta_Nx)$ for $n=2$, in agreement with Section~\ref{subsec:motivating}, and is consistent with $C_{n,1}=n!$. Table~\ref{tab:B_deficit} reports the total weight for $\beta_N=1$.

\begin{table}[H]
\centering
\caption{Total sub-mixture weight $\sum_{i\in N}p_i(x\one)$ in Example~\ref{ex:B_deficit} for $\beta_N=1$.}
\label{tab:B_deficit}
\renewcommand{\arraystretch}{1.25}
\begin{tabular}{ccccccc}
\hline
$n$ & $\alpha$ & $x=0.1$ & $x=1$ & $x=10$ & $x=100$ & $C_{n,\alpha}$\\
\hline
$2$ & $1$ & $0.9524$ & $0.6667$ & $0.1667$ & $1.96\times10^{-2}$ & $2$\\
$2$ & $2$ & $0.9286$ & $0.5185$ & $0.0394$ & $5.73\times10^{-4}$ & $6$\\
$3$ & $1$ & $0.9217$ & $0.5000$ & $0.0385$ & $5.71\times10^{-4}$ & $6$\\
$3$ & $2$ & $0.8868$ & $0.3368$ & $0.0031$ & $8.00\times10^{-7}$ & $90$\\
\hline
\end{tabular}
\end{table}

The contrast with Example~\ref{ex:B_regional} is instructive. There, every risk component is affected by an exponential risk factor confined to its own class, $\Lambda_I^{\pi}(\xx)>0$ for every gamma risk factor, and the weights converge to the strictly positive limits of Proposition~\ref{prop:GammaPareto_RV}. Here, $\B_{-}=\varnothing$: the only exponential risk factor is assigned to the leading component, so that $\Lambda_{\{j\}}^{\pi}(x\one)=0$ for $j\neq\pi(1)$, the cones $\Omega^{\pi}(x\one)=\{g_{\pi(1)}/(1+\beta_Nx)\ge g_{\pi(2)}\ge\cdots\ge g_{\pi(n)}\}$ shrink as $x$ grows, and every weight vanishes. The size of the sub-mixture deficit in the joint tail is therefore a direct reflection of the extremal dependence structure described in Section~\ref{subsec:extremal}.
\end{example}

\end{appendices}
\end{document}